\documentclass[aps,prl,twocolumn,superscriptaddress, nobibnotes]{revtex4-1}
\usepackage{amsmath}
\usepackage{bbold}
\usepackage{mathtools}
\usepackage{enumerate}
\usepackage{graphicx} 
\usepackage[caption=false]{subfig}
\usepackage{xcolor}
\usepackage{hyperref}
\hypersetup{
 breaklinks=true,
 pdfnewwindow=true,
 colorlinks=true,
 linkcolor=magenta,
 citecolor=cyan,
 filecolor=blue,
 urlcolor=cyan
}

\begin{document}

\title{Minimal-backaction work statistics of coherent engines}

\author{Milton Aguilar}
    \thanks{Both authors contributed equally to this work. Email us at \href{mailto:maguilar@itp1.uni-stuttgart.de}{Milton Aguilar} or \href{mailto:franklin.rodrigues@physik.uni-freiburg.de}{Franklin L. S. Rodrigues}.}    \affiliation{Institute for Theoretical Physics I, University of Stuttgart, D-70550 Stuttgart, Germany}

\author{Franklin L. S. Rodrigues}
    \thanks{Both authors contributed equally to this work. Email us at \href{mailto:maguilar@itp1.uni-stuttgart.de}{Milton Aguilar} or \href{mailto:franklin.rodrigues@physik.uni-freiburg.de}{Franklin L. S. Rodrigues}.}
    \affiliation{Physikalisches Institut, Albert-Ludwigs-Universit\"at Freiburg, Hermann-Herder-Stra\ss e 3, D-79104 Freiburg, Germany}
    \affiliation{EUCOR Centre for Quantum Science and Quantum Computing, Albert-Ludwigs-Universit\"at Freiburg, Hermann-Herder-Stra\ss e 3, D-79104 Freiburg, Germany}

\author{Eric Lutz}
    \affiliation{Institute for Theoretical Physics I, University of Stuttgart, D-70550 Stuttgart, Germany}

\begin{abstract}
   Determining the work statistics of quantum engines is challenging  due to measurement backaction.
 We here show that a dynamic Bayesian network-based measurement scheme, which preserves quantum coherence within an engine cycle, is minimally invasive, in the sense that the averaged measured state over one  cycle exactly coincides with the unmeasured state. It therefore provides a general framework to investigate  energy exchange statistics in quantum machines. This stands in contrast to the standard two-point measurement protocol, whose backaction can be so strong that it generally fails to reproduce the average work output of a coherent motor. It may even alter its mode of operation, causing it to cease functioning as an engine under observation. We further demonstrate that recently proposed universal fluctuation bounds do not necessarily apply to coherent machines. \end{abstract}

\maketitle

Microscopic machines are subject to thermal \cite{sch03,kay07,han09,kas17} and, at low enough temperatures, to additional quantum fluctuations \cite{kos14,ben17,mye22,can24}. These quantum fluctuations arise from random transitions between discrete energy levels during the machine's evolution, and give rise to distinctly nonclassical features \cite{mukamel_2009,talkner_2011}. As a result, common thermodynamic figures of merit, such as power output and efficiency \cite{callen_1985}, become stochastic variables. Because the presence of fluctuations strongly influences the performance of small-scale engines \cite{vdbroeck_2014,esposito_2015,seifert_2018,ryabov_2018,krishnamurthy_2019,hol21}, from biological to synthetic motors  \cite{sch03,kay07,han09,kas17,kos14,ben17,mye22,can24}, a detailed characterization of their random properties is essential. However, whereas  the fluctuation statistics of classical microscopic engines may be experimentally determined without affecting their operation \cite{bli12,mar15,kri16,mar17}, this is generally not possible in the quantum regime, where measurement backaction inevitably perturbs the system \cite{jacobs_2014}. A central issue is therefore to probe the fluctuations of a quantum machine with minimal disturbance.

Quantum coherence, encoded in the off-diagonal density matrix elements, is a fundamental attribute of nonclassical thermodynamic transformations \cite{plenio_2017}. It is, however, a fragile resource that can be readily destroyed by external perturbations, making the investigation of energy fluctuations in coherent machines   inherently difficult \cite{acin_2017,wu19,sampaio_2019,paternostro_2020,trombettoni_2021,imparato_2024,skrzypczyk_2025}. The standard two-point measurement scheme  \cite{mukamel_2009,talkner_2011}   used to evaluate the quantum statistics of energy exchanges  relies on projective energy measurements performed at the beginning and at the end of a  process \cite{hanggi_2007}. Owing to its projective nature, and the associated measurement backaction, this scheme fails to properly account for quantum coherence in the energy basis. It is thus limited to the analysis  of fluctuations in incoherent motors \cite{den20,jia21,den21,lutz_2021,fei22,serra_2024}. A prominent alternative  is  the framework of dynamic Bayesian networks which provides a general method to analyze conditional dependencies among time-dependent random quantities, from engineering and robotics to finance and biological networks \cite{neapolitan_2004,darwiche_2009}. In the quantum setting, this approach enables the inference of  energy fluctuations from indirect measurements through conditional probabilities evaluated using Bayes' rule \cite{lutz_2020,par20,mic20a,str20,zha22,lutz_2023,lutz_2024}. In contrast to the two-point measurement protocol, Bayesian networks fully capture quantum coherence throughout the evolution \cite{lutz_2020,par20,mic20a,str20,zha22,lutz_2023,lutz_2024}.

In this article, we investigate the energy exchange statistics in a generalized finite-time quantum Otto qudit engine in which coherence is dynamically generated along the cycle \cite{kos17}. This quantum machine can be viewed as a multidimensional extension of recently realized qubit-based engines \cite{ass19,pet19,hou25}. We establish that the Bayesian network approach is minimally invasive, in the sense that the average  state of the engine after a round of measurements is identical to the unmeasured state. The Bayesian network scheme therefore induces minimal measurement backaction, in sharp contrast to the two-point measurement technique. We  further show that the discrepancy between the two methods is directly linked to the presence of quantum coherence in the state of the working substance. Remarkably, the backaction associated with the two-point measurement scheme can be so strong that it qualitatively alters the operation regime of the otherwise unobserved quantum machine, driving a transition from engine to heater or accelerator. Finally, we demonstrate that recently proposed "universal" bounds on engine fluctuations, derived for quasistatic Otto cycles \cite{agarwalla_2021_2,agarwalla_2021,das23,moh23,moh23a,watanabe_2025}, can be violated in the presence of coherence.

\paragraph{Generalized coherent Otto cycle.}
We consider a quantum spin engine whose working substance $S$  is described by the density operator $\rho_i$ and  a time-dependent Hamiltonian $H_{i} (t) = \omega_{i} (t) S_{z} + g_{i}(t) S_{x}$, where $S_{x,y,z}$ are the spatial components of the $d_{S}$-dimensional spin operator, and  the index $i$ labels the four branches of the cycle. The function $\omega_{i}(t)$ controls the energy spacing of $H_{i}$, while  $g_i(t)$ modulates the amount of coherence in the basis of $S_{z}$, since $[S_{z} , S_{x}] \neq 0$.  The engine performs a finite-time quantum Otto cycle during which the working medium is repeatedly and weakly coupled \cite{davies_1974} to a hot (cold) heat reservoir at temperature $T_{\mathrm{h}}$ ($T_{\mathrm{c}}$). To make system independent statements, we characterize the bath interaction  with a generalized amplitude damping map $\mathcal{T}_{t}$ that leads to full thermalization in the infinite-time limit  \cite{nie10, srikanth_2024}. 

The four branches of the cycle are  as follows \cite{kos17}: (1) \textit{Isentropic compression}: the system $S$ undergoes unitary evolution  for a  time $t_{\mathrm{k}}$,  from state $\rho_{1}$ to state $\rho_{2} = U_{\mathrm{k}} (t_{\mathrm{k}}) \rho_{1} U_{\mathrm{k}}^{\dagger} ( t_{\mathrm{k}})$, where $U_{\mathrm{k}}$ is the time evolution operator generated by $H_{\mathrm{k}}$. The control parameters   $\omega_{\mathrm{k}}$ and $g_{\mathrm{k}}$ are chosen such that $\omega_{\mathrm{k}} (0) = \omega_{\mathrm{c}}$ and $\omega_{\mathrm{k}} (t_{\mathrm{k}}) = \omega_{\mathrm{h}}$ ($> \omega_{\mathrm{c}}$), and $g_{\mathrm{k}} (0) = g_{\mathrm{k}} (t_{\mathrm{k}}) = 0$.
    (2) \textit{Hot  isochore}: the working substance $S$ is coupled to the hot bath for a finite time $t_{\mathrm{h}}$, leading to the state $\rho_{3} = \mathcal{T}_{t_{\mathrm{h}}}^{\mathrm{h}} (\rho_{2})$. In the infinite-time limit, the system asymptotically thermalizes to the Gibbs state $\mathcal{T}_{\infty}^{\mathrm{h}} (\rho_{2}) = \text{exp} [- H_{\mathrm{k}} (t_{\mathrm{k}}) / T_{\mathrm{h}}] / \text{tr} \{ \text{exp} [- H_{\mathrm{k}} (t_{\mathrm{k}}) / T_{\mathrm{h}}] \}$.
     (3) \textit{Isentropic expansion}: during unitary evolution for time $t_{\mathrm{e}}$, the state $\rho_{3}$ transitions to state $\rho_{4} = U_{\mathrm{e}} (t_{\mathrm{e}}) \rho_{3} U_{\mathrm{e}}^{\dagger} (t_{\mathrm{e}})$, where $U_{\mathrm{e}}$ is generated by the Hamiltonian $H_{\mathrm{e}}$. The parameters  $\omega_{\mathrm{e}}$ and $g_{\mathrm{e}}$ here satisfy $\omega_{\mathrm{e}} (0) = \omega_{\mathrm{h}}$ and $\omega_{\mathrm{e}} (t_{\mathrm{e}}) = \omega_{\mathrm{c}}$, as well as $g_{\mathrm{e}} (0) = g_{\mathrm{e}} (t_{\mathrm{e}}) = 0$.
    (4) \textit{Cold isochore}: in the last stroke,   the system interacts with  the cold reservoir for a finite time $t_{\mathrm{c}}$, so that the final state is given by $\rho_{5} = \mathcal{T}_{t_{\mathrm{c}}}^{\mathrm{c}} (\rho_{4})$. In this case, the asymptotic equilibrium state reads $\mathcal{T}_{\infty}^{\mathrm{c}} (\rho_{4}) = \text{exp} [- H_{\mathrm{e}} (t_{\mathrm{e}}) / T_{\mathrm{c}}] / \text{tr} \{ \text{exp} [- H_{\mathrm{e}} (t_{\mathrm{e}}) / T_{\mathrm{c}}] \}$.

One complete engine cycle is hence  described by   the map  $\Lambda = \mathcal{T}_{t_{\mathrm{c}}}^{\mathrm{c}} \mathcal{U}_{\mathrm{e}} \mathcal{T}_{t_{\mathrm{h}}}^{\mathrm{h}} \mathcal{U}_{\mathrm{k}}$, where $\mathcal{U}_{i} (\cdot) = U_{i} (t_{i}) \cdot U_{i}^{\dagger} (t_{i})$ with a total duration of $\tau = t_{\mathrm{k}} + t_{\mathrm{h}} + t_{\mathrm{e}} + t_{\mathrm{c}}$. After an initial  transient, the motor will reach a limit-cycle regime for the state $\rho_1$ characterized by the fixed point of the map, $ \Lambda (\tilde{\rho}_{1}) = \tilde{\rho}_{1}$ \cite{com}. In general, this state will  exhibit quantum coherence in the   basis of $S_{z}$ for nonzero transverse drivings  $g_i(t)$ and finite bath-coupling times $t_i$. Such coherence impacts the thermodynamic performance of the machine. The  fluctuation analysis presented below will thus be based on these limit-cycle states.

We begin by recalling the thermodynamics of the unobserved, that is, unperturbed, quantum Otto engine \cite{kos17}.
During the  hot and cold isochoric strokes, the system Hamiltonian  is held constant. As a consequence, no work is performed, and only heat is exchanged with the two thermal baths. In both cases, the exchanged heat is given by the variation of the mean (internal) energy  of the working substance:  $Q_{\mathrm{h}} = \omega_{\mathrm{h}} \text{tr} \{ S_{z} [  \mathcal{T}_{t_{\mathrm{h}}}^{\mathrm{h}} \mathcal{U}_{\mathrm{k}} (\tilde{\rho}_1) - \mathcal{U}_{\mathrm{k}} (\tilde{\rho}_1)] \}$ and $ Q_{\mathrm{c}} = - \omega_{\textrm{c}} \text{tr} \{ S_{z} [\mathcal{U}_{\mathrm{e}} \mathcal{T}_{t_{\mathrm{h}}}^{\mathrm{h}} \mathcal{U}_{\mathrm{k}} (\tilde{\rho}_1) - \tilde{\rho}_1] \}$. The total work then follows from the first law of thermodynamics as $W = Q_{\mathrm{h}} + Q_{\mathrm{c}}$ \cite{kos17}. Engine operation  is characterized by a positive work output, $W > 0$, together with the condition  that heat is absorbed from the hot bath, $Q_{\textrm{h}} > 0$, and released into the cold one, $Q_{\textrm{c}} < 0$.

\paragraph{Fluctuation statistics of a coherent engine.}
The  distribution $P (w)$ of the stochastic  work output of a quantum  machine  can be obtained by sequentially measuring the working substance at the four corners of the cycle. Its general form is \cite{den20,jia21,den21,lutz_2021,fei22,serra_2024}
\begin{equation}
    \begin{aligned}
        P (w) & = \sum_{j,l,m,n} \delta [w + (e_{\mathrm{k}}^{l} - e_{\mathrm{e}}^{j}) + (e_{\mathrm{e}}^{n} - e_{\mathrm{k}}^{m})] \\
        & \times p_{w} (e_{\mathrm{e}}^{j} , e_{\mathrm{k}}^{l} , e_{\mathrm{k}}^{m}, e_{\mathrm{e}}^{n}),
    \end{aligned}
    \label{eq:wGeneral}
\end{equation}
where $e_{\mathrm{k} , \mathrm{e}}^{j}$ is the $j$-th eigenvalue of the Hamiltonian $H_{\mathrm{k} , \mathrm{e}} (t_{\mathrm{k} , \mathrm{e}})$ and $p_{w} (e_{\mathrm{e}}^{j} , e_{\mathrm{k}}^{l} , e_{\mathrm{k}}^{m}, e_{\mathrm{e}}^{n})$ denotes the joint probability of the four outcomes \cite{den20,jia21,den21,lutz_2021,fei22,serra_2024}. The latter quantity depends on the chosen measurement protocol, and hence on  the associated measurement backaction.

To illustrate the difference between the two-point measurement (TPM) scheme \cite{mukamel_2009,talkner_2011} and the dynamic Bayesian network (DBN) approach \cite{neapolitan_2004,darwiche_2009}, we consider  an arbitrary evolution taking a state $\rho_{\textrm{i}}$ of a  system to another state $\rho_{\textrm{f}} = \mathrm{M} (\rho_{\textrm{i}})$, with $\mathrm{M}$ a CPTP map \cite{nie10}, while also changing its Hamiltonian from $H_{\textrm{i}}$ to $H_{\textrm{f}}$.  The  distribution of the random  energy change $\Delta u$ of the system during this process is  $P (\Delta u ) = \sum_{j,k} \delta [\Delta u - (e_{\textrm{f}}^{k} - e_{\textrm{i}}^{j})] \, p (e_{\textrm{i}}^{j} , e_{\textrm{f}}^{k})$. Within the TPM scheme, the joint probability $p (e_{\textrm{i}}^{j} , e_{\textrm{f}}^{k})$ is obtained by first performing a projective measurement in the eigenbasis of $H_{\textrm{i}}$ on $\rho_{\textrm{i}}$, and then measuring the evolved projected state in the eigenbasis of $H_{\textrm{f}}$   
 \cite{mukamel_2009,talkner_2011}:
    $p_{\textrm{TPM}} (e_{\textrm{i}}^{j} , e_{\textrm{f}}^{k}) = \text{tr} [ \Pi_{\textrm{f}}^{k} \, \mathrm{M} ( \Pi_{\textrm{i}}^{j} \rho_{\textrm{i}} \Pi_{\textrm{i}}^{j}) ]$,
where $\Pi_{\textrm{i}, \textrm{f}}^{j}$ is the eigenprojector corresponding to $e_{\textrm{i},\textrm{f}}^{j}$. The first measurement collapses $\rho_{\textrm{i}}$ into an eigenspace of $H_{\textrm{i}}$, and therefore destroys all its coherences in that basis. The same happens with the second projective measurement. 

On the other hand,  the dynamic Bayesian network method is based  on projective measurements in the eigenbasis of $\rho_{\textrm{i} , \textrm{f}}$, instead of that of $H_{\textrm{i} , \textrm{f}}$, hence preserving initial and final coherences in the energy basis \cite{lutz_2020,par20,mic20a,str20,zha22,lutz_2023,lutz_2024}. The joint probability $p_{\textrm{DBN}} (e_{\textrm{i}}^{j} , e_{\textrm{f}}^{k})$ of the unmeasured energy eigenvalues is then obtained using Bayesian inference.   Let us concretely write the eigendecomposition of the initial and final states as $\rho_{\textrm{i} , \textrm{f}} = \sum_{\alpha} \lambda_{\textrm{i} , \textrm{f}}^{\alpha} \mathrm{P}_{\textrm{i} , \textrm{f}}^{\alpha}$ where  $\mathrm{P}_{\textrm{i} , \textrm{f}}^{\alpha}$ is the eigenprojector corresponding to the eigenvalue $\lambda_{\textrm{i} , \textrm{f}}^{\alpha}$. By now performing the first projective measurement on the complete measurement ensemble, dividing it into two halves, time evolving one of them and then projectively measuring it, the joint  distribution of measurement outcomes is $p (\lambda_{\mathrm{i}}^{\alpha} , \lambda_{\mathrm{f}}^{\beta}) = \text{tr} [ \mathrm{P}_{\textrm{f}}^{\beta} \, \mathrm{M} ( \mathrm{P}_{\textrm{i}}^{\alpha} \rho_{\textrm{i}} \mathrm{P}_{\textrm{i}}^{\alpha})]$. The next step is to evaluate the posterior conditional probabilities for each Hamiltonian eigenvalue using the standard Born rule, $p (e_{\mathrm{i}}^{j} | \lambda_{\mathrm{i}}^{\alpha}) = \text{tr} [\Pi_{\textrm{i}}^{j} \mathrm{P}_{\textrm{i}}^{\alpha} \rho_{\textrm{i}} \mathrm{P}_{\textrm{i}}^{\alpha} / \text{tr} (\mathrm{P}_{\textrm{i}}^{\alpha} \rho_{\textrm{i}})]$ and $p (e_{\mathrm{f}}^{k} | \lambda_{\mathrm{f}}^{\beta}) = \text{tr} [\Pi_{\textrm{f}}^{k} \mathrm{P}_{\textrm{f}}^{\beta} \rho_{\textrm{f}} \mathrm{P}_{\textrm{f}}^{\beta} / \text{tr} (P_{\textrm{f}}^{\beta} \rho_{\textrm{f}})]$. Summing over all possible measurement outcomes, one finds \cite{lutz_2020,par20,mic20a,str20,zha22,lutz_2023,lutz_2024}, 
    $p_{\textrm{DBN}} (e_{\textrm{i}}^{j} , e_{\textrm{f}}^{k}) = \sum_{\alpha , \beta} p (e_{\mathrm{i}}^{j} | \lambda_{\mathrm{i}}^{\alpha}) \, p (e_{\mathrm{f}}^{k} | \lambda_{\mathrm{f}}^{\beta}) \, p (\lambda_{\mathrm{i}}^{\alpha} , \lambda_{\mathrm{f}}^{\beta})
$. The two joint distributions, $p_{\textrm{TPM}}$ and  $p_{\textrm{DBN}}$, coincide in the absence of quantum coherence in the energy basis, yielding identical energy statistics. However, because the states $\rho_{\textrm{i} , \textrm{f}}$ are diagonal in their eigenbases, measurement backaction is strongly reduced in the DBN protocol in the presence of energy coherences, compared to the TPM approach. Importantly, the Bayesian network distribution $p_{\textrm{DBN}} (e_{\textrm{i}}^{j} , e_{\textrm{f}}^{k}) $ is accessible experimentally \cite{mic20a}.

\begin{figure}[!t]
    \centering
    \includegraphics[width=1\linewidth]{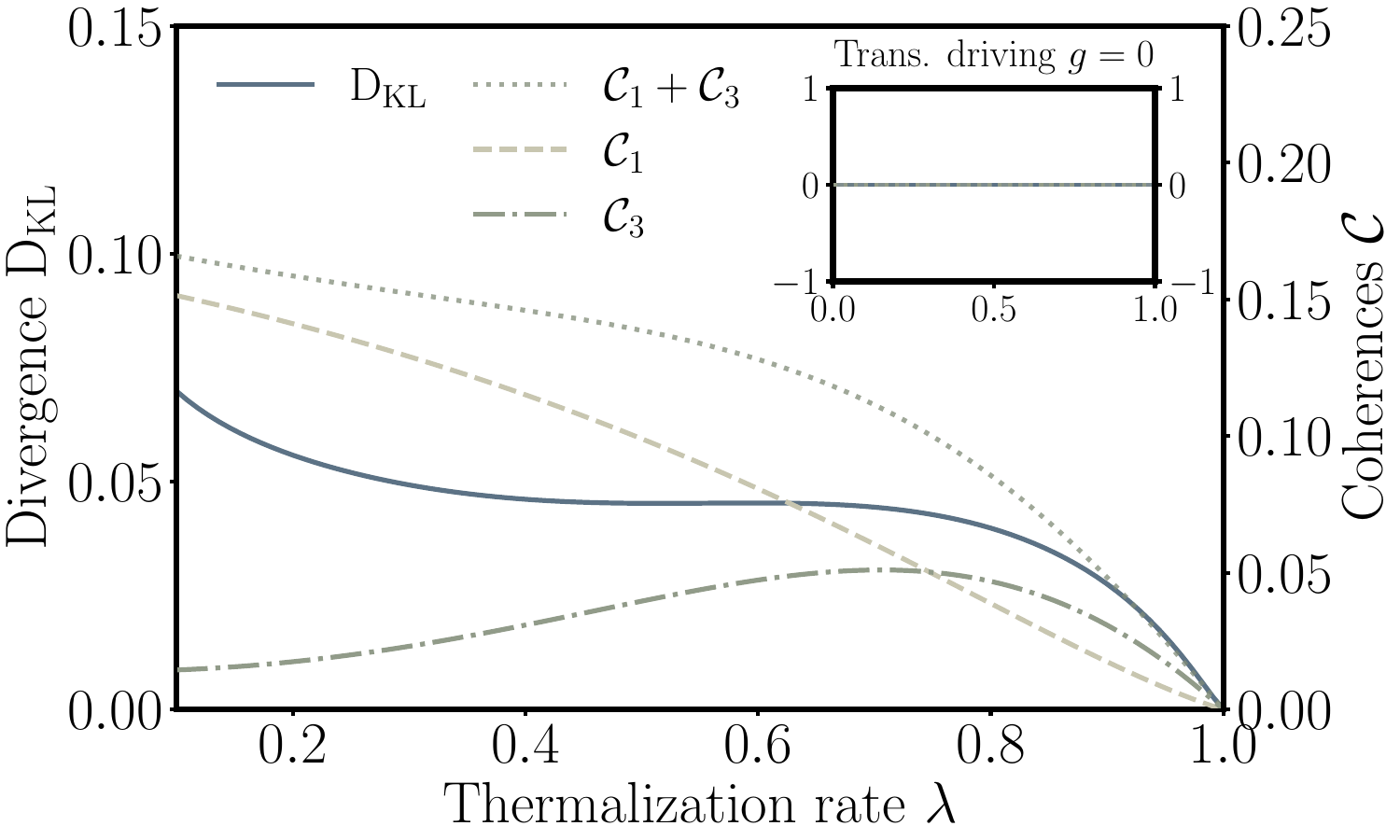}
    \caption{Entropic distance, $D_\textrm{KL} (P_{\textrm{DBN}} || P_{\textrm{TPM}})$, between the  work distributions of the dynamic Bayesian network (DBN)  approach and the two-point measurement (TPM) scheme versus the relative entropies of coherence, $\mathcal{C}_{1,3}$, after the two isochores in  a three-level   quantum  Otto cycle. The largest difference is observed for maximal coherence.  It vanishes for zero transverse driving strength $g=0$ (inset). Parameters  are $\omega_{\mathrm{h}} = 10$, $T_{\mathrm{h}} = 14$, $\omega_{\mathrm{c}} = 0.5$, $T_{\mathrm{c}} = 0.1$, $g = 9$, $\lambda_{\mathrm{c}} = \lambda_{\mathrm{h}} = \lambda$.} 
    \label{fig:cor}
\end{figure}

The joint probability $p_{w} (e_{\mathrm{e}}^{j} , e_{\mathrm{k}}^{l} , e_{\mathrm{k}}^{m}, e_{\mathrm{e}}^{n})$ in Eq.~\eqref{eq:wGeneral} can be derived for the quantum Otto engine  in a similar manner by extending the previous considerations to the four corners of the cycle. The explicit (and lengthy) expressions for the work output distribution $P(w)$ for the two measurement schemes are given in the Supplemental Material. We here indicate the corresponding average work outputs $\langle w \rangle = \int dw \, w \, P(w)$. We have
\begin{eqnarray}
        \langle w \rangle_{\mathrm{TPM}} & =& \omega_{\mathrm{h}} \text{tr} \{ {J}_{z} [  \mathrm{D}_{\mathrm{k}} \mathcal{T}_{t_{\mathrm{h}}}^{\mathrm{h}} \mathcal{U}_{\mathrm{k}} \mathrm{D}_{\mathrm{e}} (\tilde{\rho}_{1}) - \mathcal{U}_{\mathrm{k}} \mathrm{D}_{\mathrm{e}} (\tilde{\rho}_{1})] \} \nonumber \\
        && - \omega_{\mathrm{c}} \text{tr} \{ {J}_{z} [ \mathcal{U}_{\mathrm{e}} \mathrm{D}_{\mathrm{k}} \mathcal{T}_{t_{\mathrm{h}}}^{\mathrm{h}} \mathcal{U}_{\mathrm{k}} \mathrm{D}_{\mathrm{e}} (\tilde{\rho}_{1}) - \tilde{\rho}_{1}] \},\\
         \langle w \rangle_{\mathrm{DBN}} & =& \omega_{\mathrm{h}} \text{tr} \{ {J}_{z} [  \mathcal{T}_{t_{\mathrm{h}}}^{\mathrm{h}} \mathcal{U}_{\mathrm{k}} (\tilde{\rho}_{1}) - \mathcal{U}_{\mathrm{k}} (\tilde{\rho}_{1})] \} \nonumber \\
        && - \omega_{\mathrm{c}} \text{tr} \{ {J}_{z} [ \mathcal{U}_{\mathrm{e}} \mathcal{T}_{t_{\mathrm{h}}}^{\mathrm{h}} \mathcal{U}_{\mathrm{k}} (\tilde{\rho}_{1}) - \tilde{\rho}_{1}] \},
    \label{eq:wDBN}
\end{eqnarray}
where $\mathrm{D}_{\textrm{i}} (\cdot) = \sum_{j} \Pi_{\textrm{i}}^{j} \cdot \Pi_{\textrm{i}}^{j}$ is the dephasing map \cite{nie10} (for the TPM result, we have additionally used $[ \mathrm{D}_{\mathrm{k}} , \mathcal{T}_{t_{\mathrm{h}}}^{\mathrm{h}} ] = 0$).

Remarkably, the average work, $\langle w \rangle_{\mathrm{DBN}}$ in Eq.~(3), obtained via Bayesian networks always  agrees exactly with the unmeasured result $W$. By contrast, the two-point measurement expression,  $\langle w \rangle_{\mathrm{TPM}}$ in  Eq.~(2), differs from it in general due to measurement backaction. A comparison of Eqs.~(2) and (3) reveals that this discrepancy originates from  the presence of quantum coherence during the cycle: $\langle w \rangle_{\mathrm{TPM}}= \langle w \rangle_{\mathrm{DBN}}=W$ only if the states of the working substance after the hot and cold isochores are diagonal in the eigenbasis of their respective Hamiltonians, that is,   $\tilde{\rho}_{1} = \mathrm{D}_{\mathrm{e}} (\tilde{\rho}_{1})$ after the cold isochore and  $\mathcal{T}_{t_{\mathrm{h}}}^{\mathrm{h}} \mathcal{U}_{\mathrm{k}} \mathrm{D}_{\mathrm{e}} (\tilde{\rho}_{1}) = \mathrm{D}_{\mathrm{k}} \mathcal{T}_{t_{\mathrm{h}}}^{\mathrm{h}} \mathcal{U}_{\mathrm{k}} \mathrm{D}_{\mathrm{e}} (\tilde{\rho}_{1})$ after the hot isochore. For the specific Hamiltonian considered here, this is achieved either in the absence of the transverse driving ($g_i = 0$), or when the system fully thermalizes with the baths ($t_{\mathrm{h} , \mathrm{c}} \to \infty$). In these cases  the two work distributions (1) also coincide, $P_{\mathrm{TPM}}(w) = P_{\mathrm{DBN}}(w)$. For arbitrary system Hamiltonians $H_i$, and finite thermalization times, these conditions can be generically formulated as (Supplemental Material)
\begin{equation}
   \!\!  \!   \mathcal{U}_{\mathrm{k}} \mathrm{D}_{\mathrm{e}} (\rho) \!= \mathrm{D}_{\mathrm{k}} \mathcal{U}_{\mathrm{k}} \mathrm{D}_{\mathrm{e}} (\rho)  \text{ and }  \mathcal{U}_{\mathrm{e}} \mathrm{D}_{\mathrm{k}} ( \rho) \!= \mathrm{D}_{\mathrm{e}} \mathcal{U}_{\mathrm{e}} \mathrm{D}_{\mathrm{k}} (\rho)  \!  \!        \label{eq:conditionTPM}
\end{equation}
since the output of the generalized amplitude damping map $\mathcal{T}_{t}$ is diagonal only if the input is already diagonal. Equation \eqref{eq:conditionTPM} is satisfied whenever the evolution operators $\mathcal{U}_{\mathrm{k,e}}$ do not induce coherences on incoherent states. It is worth noting that dynamic Bayesian networks are different from so-called collective measurements which have been shown to reduce measurement backaction  \cite{acin_2017,wu19}. Whereas the measured average work in that scheme approaches the unmeasured value, after proper parameter optimization, it does not reach it in general.  

To characterize the difference between the two work distributions, we introduce their entropic distance (or Kullback-Leibler divergence), $D_\textrm{KL} (P_{\textrm{DBN}} || P_{\textrm{TPM}})=\int dw\, P_{\textrm{DBN}}(w) \ln[P_{\textrm{DBN}}(w)/P_{\textrm{TPM}}(w)]$ \cite{cover_2005}. We furthermore evaluate the amount of quantum coherence in the  state of the system with  the relative entropy of coherence, $\mathcal{C}(\tilde \rho)=\text{tr}\{\text{D}(\tilde{\rho}) \{ \ln[\text{D}(\tilde{\rho})]-\ln(\tilde{\rho})\} \}$ \cite{plenio_2017}. Figure~\ref{fig:cor} shows $D_\textrm{KL} (P_{\textrm{DBN}} || P_{\textrm{TPM}})$, as well as $\mathcal{C}(\tilde \rho_1)$ and $\mathcal{C}(\tilde \rho_3)$, the coherences at the end of the two isochores, for a three-level system with frequencies $\omega_{\mathrm{k} , \mathrm{e}} (t) = (\omega_{\mathrm{h} , \mathrm{c}} - \omega_{\mathrm{c} , \mathrm{h}}) t / t_{\mathrm{k} , \mathrm{e}} + \omega_{\mathrm{c} , \mathrm{h}}$ and transverse drivings $g_{\mathrm{k} , \mathrm{e}} (t) = g \, \text{sin} (\pi \, t / t_{\mathrm{k} , \mathrm{e}})$, as a function the  thermalization rate $\lambda$. The latter is related to the thermalization time via $t_{\mathrm{h} , \mathrm{c}} = - \text{ln} ( 1 - \lambda_{\mathrm{h} , \mathrm{c}} )$; we set $t_{\mathrm{h}} = t_{\mathrm{c}}$ for simplicity. We observe a strong correlation between the  difference in the two work distributions and the amount of   quantum coherence generated during the cycle. In particular, the difference becomes maximal for parameter regimes in which the state of the working medium exhibits the largest degree of coherence.

\begin{figure}[!t]
    \centering
    \includegraphics[width=1\linewidth]{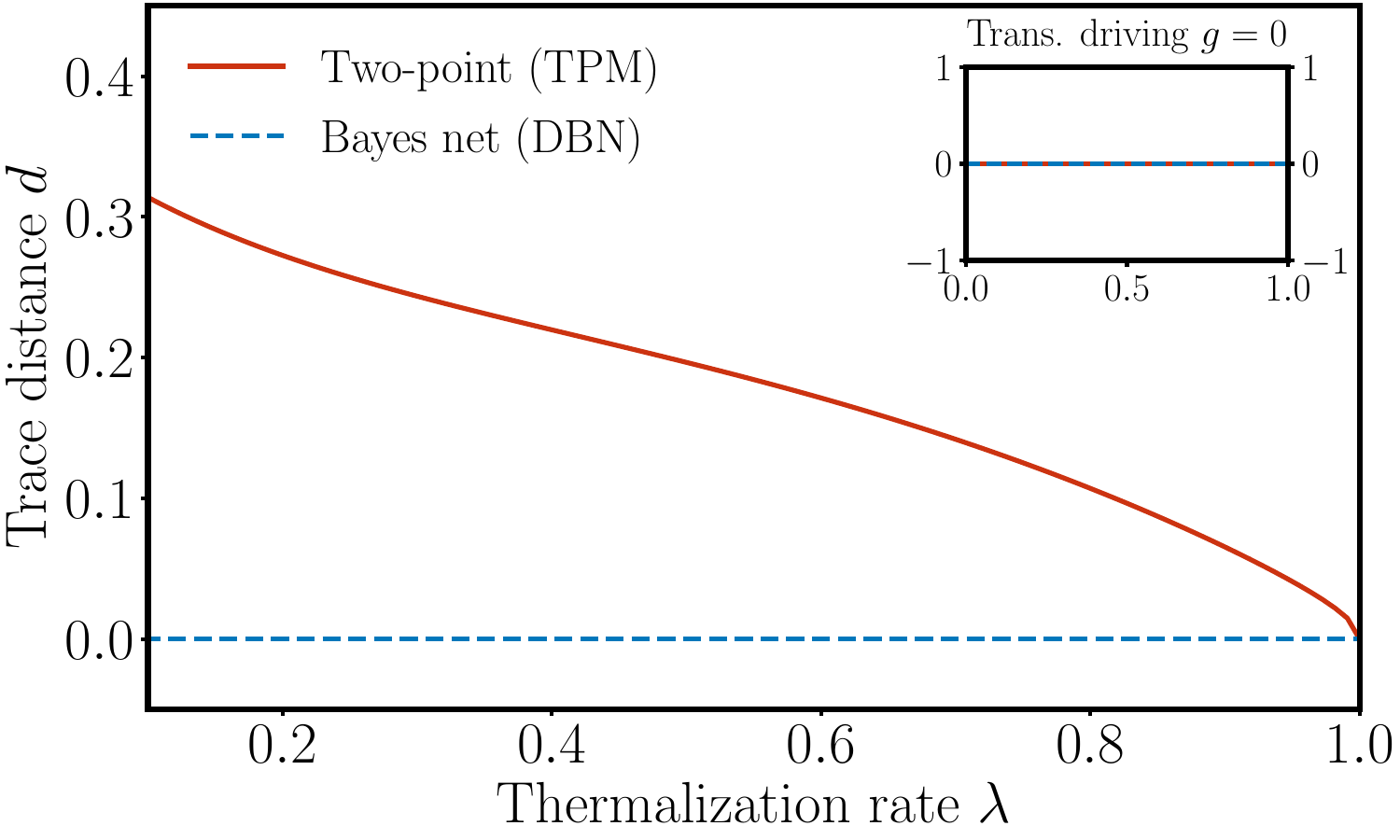}
    \caption{Trace distance, $d(\langle \rho \rangle, \tilde \rho_1)$, between the averaged measured state $\langle \rho\rangle$ and the unmeasured steady state $\tilde \rho_1$ for the two measurement protocols (TPM and DBN). It  vanishes for dynamic Bayesian networks, showing that the latter minimizes measurement backaction. Same parameters as in Fig.~1.}
    \label{fig:distance}
\end{figure}

\begin{figure}[!ht]
	\includegraphics[width=1\linewidth]{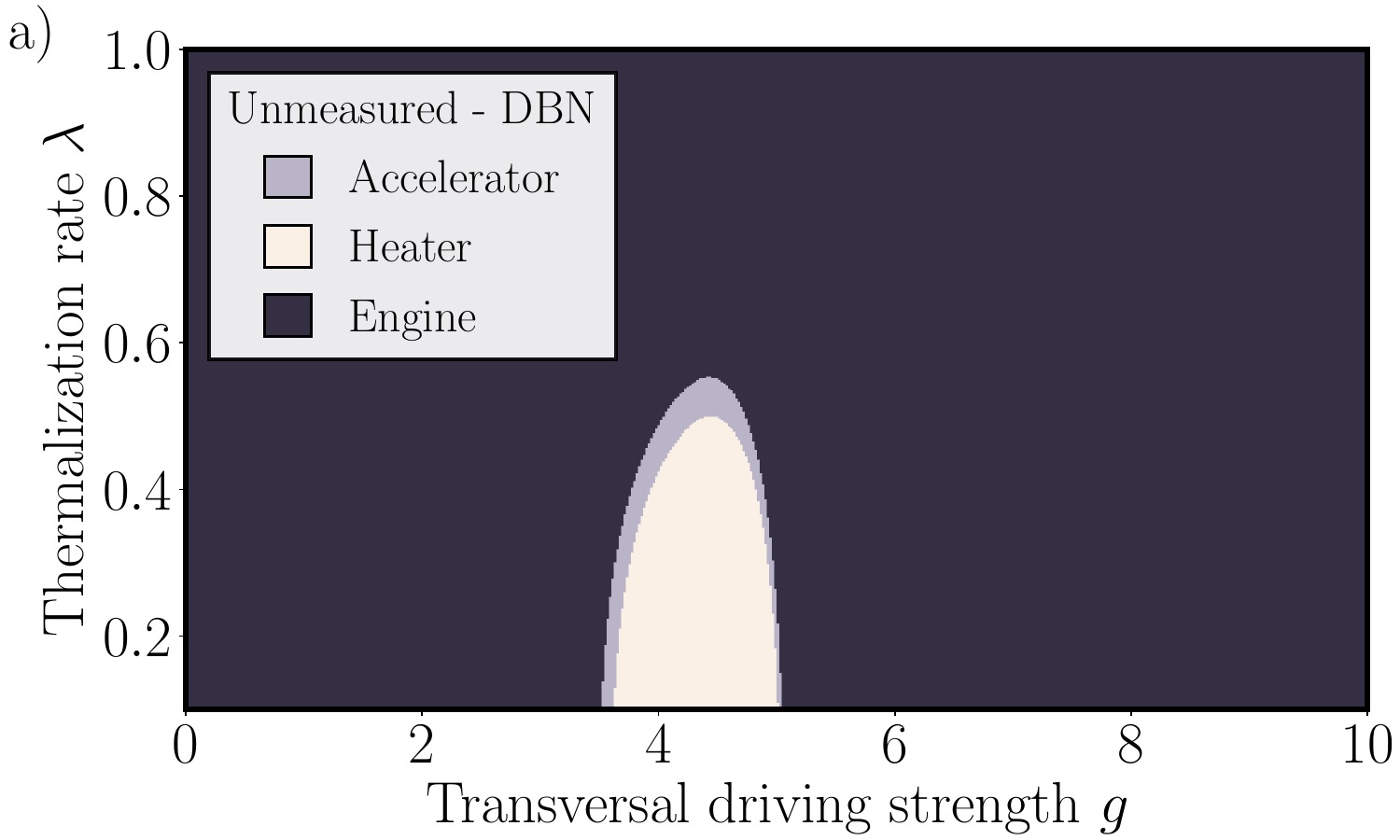}%
	
    \includegraphics[width=1\linewidth]{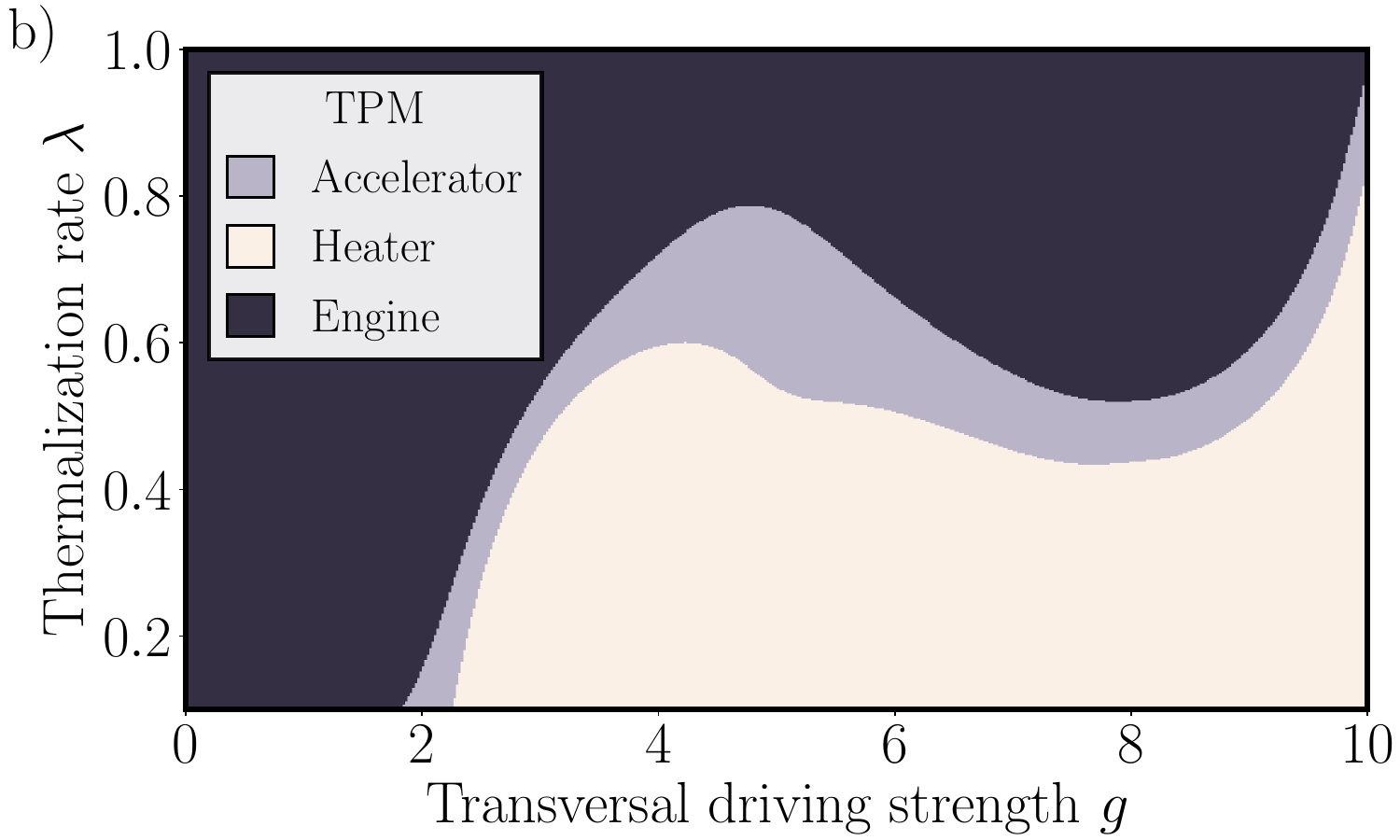}%
	\caption{Operation regimes of the quantum Otto machine for the two measurement schemes. a) Unmeasured and DBN-measured machines have identical mean work and heat, and hence display identical behavior. b) By contrast, the TPM-measured machine exhibits radically modified regimes due to measurement backaction. Same parameters as in Fig.~1.}
    \label{fig:regimes}
\end{figure}

\begin{figure}[t]
    \centering
    \includegraphics[width=1\linewidth]{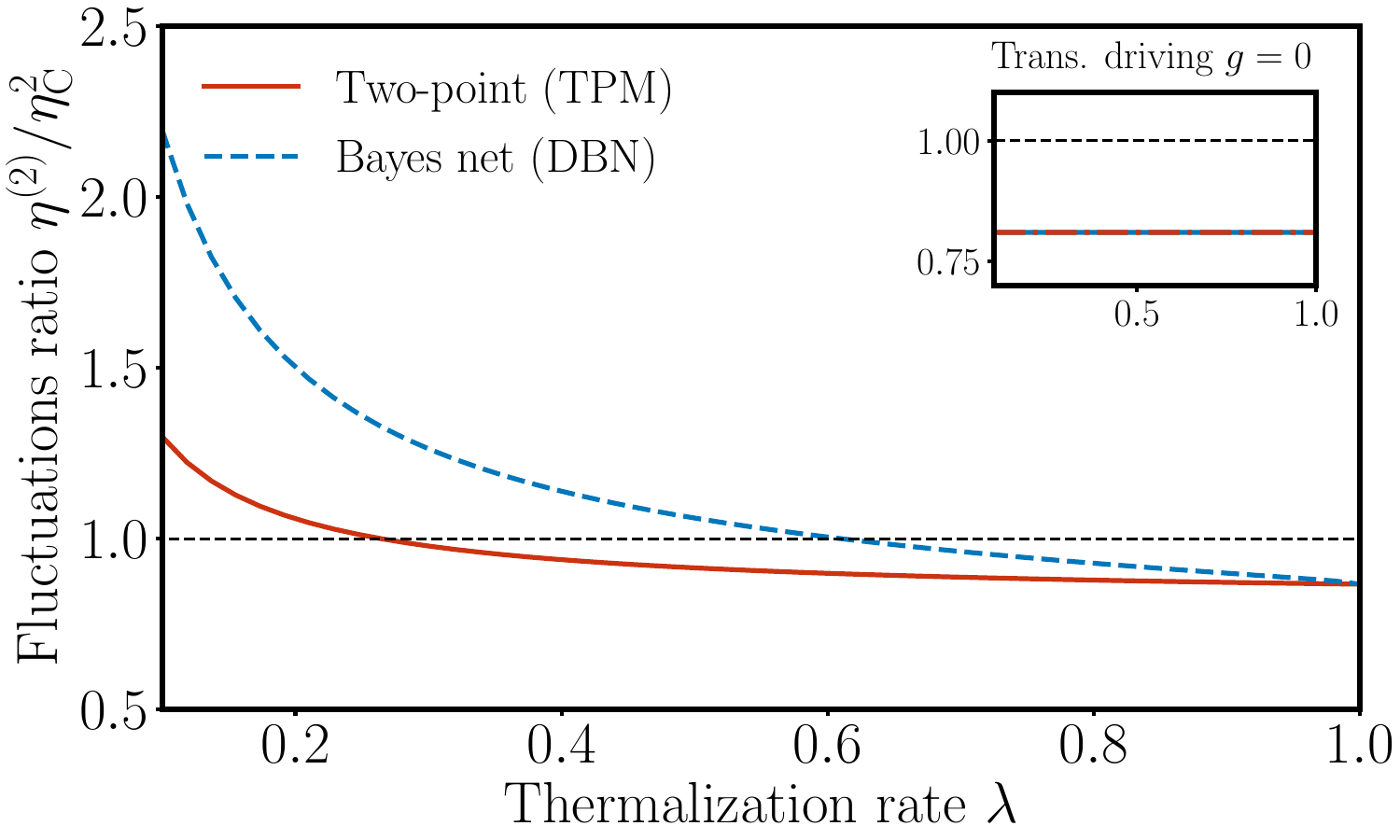}
    \caption{The fluctuation bound,  $\eta^{(2)}/ \eta_{C}^{2}\leq 1$, Eq.~(5), can be violated both for the TPM and DBN protocols owing to quantum coherence within the engine cycle for small thermalization rates $\lambda$,  in the quasistatic regime.  Parameters  are $\omega_{\mathrm{h}} = 1$, $T_{\mathrm{h}} = 1.2$, $\omega_{\mathrm{c}} = 0.85$, $T_{\mathrm{c}} = 1$, $g = 0.06$, and $\lambda_{\mathrm{c}} = \lambda_{\mathrm{h}} = \lambda$.}
    \label{fig:violation}
\end{figure}

\paragraph{Minimal measurement backaction.}
Both measurement protocols are unavoidably accompanied by measurement backaction. However, the above findings  suggest that the backaction induced by the dynamic Bayesian network scheme is significantly weaker than that of the two-point measurement approach. To quantify the disturbance induced by a measurement sequence, we compare  the unmeasured steady state $\tilde \rho_1$ with  the  averaged state obtained after a full engine cycle interspersed by five measurements, back to the initial corner of the cycle. We find $\langle \rho \rangle_{\textrm{TPM}} = \mathrm{D}_{\mathrm{e}} \mathcal{T}_{t_{\mathrm{c}}}^{\mathrm{c}} \mathrm{D}_{\mathrm{e}} \mathcal{U}_{\mathrm{e}} \mathrm{D}_{\mathrm{k}} \mathcal{T}_{t_{\mathrm{h}}}^{\mathrm{h}} \mathrm{D}_{\mathrm{k}} \mathcal{U}_{\mathrm{k}} \mathrm{D}_{\mathrm{e}} (\tilde{\rho}_{1}) \neq \tilde \rho_1$ whereas $\langle \rho \rangle_{\textrm{DBN}} = \tilde{\rho}_{1}$.
The Bayesian network approach therefore does not perturb the state of the working substance on average. In this sense, it is minimally invasive, in sharp contrast to the two-point measurement protocol. This difference arises because, in the Bayesian network scheme, measurements are performed in the eigenbasis of the density operator, in which the state is trivially diagonal. Figure \ref{fig:distance} displays the trace distance, $d(\langle \rho \rangle, \tilde \rho_1)$ \cite{nie10}, between the averaged measured state and the unmeasured state, as a function of the thermalization rate $\lambda$. These results indicate that  two-point measurement experiments become increasingly invasive as the amount of  coherence generated during the engine cycle grows for smaller $\lambda$.

Measurement backaction in the two-point measurement framework can, in fact, become so strong that it completely alters the operation regime of the machine. This behavior is illustrated in Fig.~\ref{fig:regimes} as a function  of the thermalization rate and the transverse driving strength. Depending on the parameter regime, the TPM-measured machine can run as an accelerator, that uses work to enhance the natural heat flow between the baths ($\langle w \rangle < 0$, $\langle q_{\mathrm{h}} \rangle > 0$, $\langle q_{\mathrm{c}} \rangle < 0$), or a heater, that uses work to heat up both baths ($\langle w \rangle < 0$, $\langle q_{\mathrm{h}} \rangle < 0$, $\langle q_{\mathrm{c}} \rangle < 0$) \cite{sol20}, while the unmeasured (and DBN-measured) machine functions as an engine, that uses the heat flow between the baths to output work ($\langle w \rangle > 0$, $\langle q_{\mathrm{h}} \rangle > 0$, $\langle q_{\mathrm{c}} \rangle < 0$). The difference between the two, which can be related to a modification of the first law by the TPM-scheme (Supplemental Material), is especially marked for small thermalization rates $\lambda$ and large transverse driving strengths $g$.

\paragraph{Violation of fluctuation bounds.}
Probability distributions not only give access to the mean but also to fluctuations around the mean. Recently, "universal" bounds on such fluctuations have been proposed for quasistatic quantum heat engines  in the form \cite{agarwalla_2021_2,agarwalla_2021,das23,moh23,moh23a,watanabe_2025}
\begin{equation}
    \eta^{(2)} = \frac{\sigma_{w}^{2}}{\sigma_{q_{\mathrm{h}}}^{2}} \leq \eta_{C}^{2} = \left( 1 - \frac{T_{\mathrm{c}}}{T_{\mathrm{h}}} \right)^{2},
    \label{eq:bound1}
\end{equation}
where $\sigma^2_x$ denotes the variance of the random variable $x$. Such upper bounds on the work output fluctuations are important, since they determine the stability of a thermal machine, both in the classical \cite{seifert_2018,ryabov_2018} and in the quantum \cite{lutz_2021,fei22} domains: an engine is usually considered stable, if the work output fluctuations are smaller than the mean. The explicit expressions of the variances for the two measurement schemes may  be found in the Supplemental Material.

Equation \eqref{eq:bound1} was derived assuming full thermalization and in the absence of quantum coherence \cite{agarwalla_2021_2,agarwalla_2021,das23,moh23,moh23a,watanabe_2025}. Figure \ref{fig:violation} reveals that the bound  is actually violated in the presence of quantum coherence, and this for both measurement schemes, for finite relaxation times. Coherence is seen to augment work output fluctuations. We  stress that  the value of the transverse driving  strength is here one order of magnitude smaller than the frequencies to ensure that the engine runs in the quasistatic regime.

\paragraph{Conclusions.} Quantum signatures manifest not only in the mean values of thermodynamic variables but also in their fluctuations. It is therefore crucial to develop an effective method to determine the energy exchange statistics of quantum machines with  minimal measurement backaction, to preserve coherent features.  We have shown that the dynamic Bayesian network approach  is minimally invasive, in the sense that the averaged measured state exactly coincides with the unmeasured one, unlike the two-point measurement protocol. The measurement backaction of the latter can be so strong that it generally fails  to reproduce the average work output of a coherent engine and may even dramatically alter its operation mode. In such cases, the measured motor may cease to function as a heat engine altogether.  Furthermore, we have   found that recently derived "universal" fluctuation bounds do not necessarily apply to coherent engines. Our results are not restricted to the quantum Otto cycle, and hence offer a versatile platform for both  theoretical and experimental investigations of the energy exchange statistics of coherent machines.

\textit{Acknowledgements.} Financial support from the
German Research Foundation DFG (Grant FOR 2724) and the Georg H. Endress
Foundation is acknowledged.

\clearpage
\widetext
\begin{center}
\textbf{\large Supplemental Material: Minimal-backaction work statistics of coherent engines}
\end{center}
\setcounter{equation}{0}
\setcounter{figure}{0}
\setcounter{table}{0}
\setcounter{page}{1}
\setcounter{secnumdepth}{4}
\makeatletter
\renewcommand{\theequation}{S\arabic{equation}}
\renewcommand{\thefigure}{S\arabic{figure}}
\renewcommand{\bibnumfmt}[1]{[S#1]}
\renewcommand{\citenumfont}[1]{S#1}


\renewcommand{\figurename}{Supplementary Figure}
\renewcommand{\theequation}{S\arabic{equation}}
\renewcommand{\thefigure}{S\arabic{figure}}
\renewcommand{\bibnumfmt}[1]{[S#1]}
\renewcommand{\citenumfont}[1]{S#1}

\section{Work and heat probability distributions: mean values and variances}

We begin by explaining how to compute the work and heat probability distributions using the two-point measurement (TPM) scheme and the dynamic Bayesian network (DBN) approach. Their general form is \cite{den20s,jia21s,den21s,lutz_2021s,fei22s,serra_2024s}
\begin{equation}
    P (w) = \sum_{j,l,m,n} \delta [w + (e_{\mathrm{k}}^{l} - e_{\mathrm{e}}^{j}) + (e_{\mathrm{e}}^{n} - e_{\mathrm{k}}^{m})] \, p_{w} (e_{\mathrm{e}}^{j} , e_{\mathrm{k}}^{l} , e_{\mathrm{k}}^{m}, e_{\mathrm{e}}^{n}),
    \label{eqsm:wGeneral}
\end{equation}
for the work probability distribution, and
\begin{equation}
    P (q_{\mathrm{h}}) = \sum_{l,m} \delta [q_{\mathrm{h}} - (e_{\mathrm{k}}^{m} - e_{\mathrm{k}}^{l})] \, p_{q_{\mathrm{h}}} (e_{\mathrm{k}}^{l} , e_{\mathrm{k}}^{m}),
    \label{eqsm:qhGeneral}
\end{equation}
and
\begin{equation}
    P (q_{\mathrm{c}}) = \sum_{n,r} \delta [q_{\mathrm{c}} - (e_{\mathrm{e}}^{r} - e_{\mathrm{e}}^{n}) ] \, p_{q_{\mathrm{c}}} (e_{\mathrm{e}}^{n} , e_{\mathrm{e}}^{r}),
    \label{eqsm:qcGeneral}
\end{equation}
for the heat exchanged probability distributions with the hot and cold baths, respectively. Above, $e_{\mathrm{k} , \mathrm{e}}^{j}$ is the $j$-th eigenvalue of the Hamiltonian  $H_{\mathrm{k} , \mathrm{e}} (t_{\mathrm{k} , \mathrm{e}})$ and $p_{w}$, $p_{q_{\mathrm{h}}}$ and $p_{q_{\mathrm{c}}}$ are the joint probabilities of obtaining the corresponding eigenvalues. We note that $p_{w}$, $p_{q_{\mathrm{h}}}$ and $p_{q_{\mathrm{c}}}$ can all be obtained from a single probability distribution $p (e_{\mathrm{e}}^{j} , e_{\mathrm{k}}^{l} , e_{\mathrm{k}}^{m}, e_{\mathrm{e}}^{n}, e_{\mathrm{e}}^{r})$ by computing the associated marginals:
\begin{equation}
    \begin{aligned}
        p_{w} (e_{\mathrm{e}}^{j} , e_{\mathrm{k}}^{l} , e_{\mathrm{k}}^{m}, e_{\mathrm{e}}^{n}) & = \sum_{r} p (e_{\mathrm{e}}^{j} , e_{\mathrm{k}}^{l} , e_{\mathrm{k}}^{m}, e_{\mathrm{e}}^{n}, e_{\mathrm{e}}^{r}), \\
        p_{q_{\mathrm{h}}} (e_{\mathrm{k}}^{l} , e_{\mathrm{k}}^{m}) & = \sum_{r,n,j} p (e_{\mathrm{e}}^{j} , e_{\mathrm{k}}^{l} , e_{\mathrm{k}}^{m}, e_{\mathrm{e}}^{n}, e_{\mathrm{e}}^{r}), \\
        p_{q_{\mathrm{c}}} (e_{\mathrm{e}}^{n} , e_{\mathrm{e}}^{r}) & = \sum_{j,l,m} p (e_{\mathrm{e}}^{j} , e_{\mathrm{k}}^{l} , e_{\mathrm{k}}^{m}, e_{\mathrm{e}}^{n}, e_{\mathrm{e}}^{r}).
    \end{aligned}
\end{equation}
Operationally, $p (e_{\mathrm{e}}^{j} , e_{\mathrm{k}}^{l} , e_{\mathrm{k}}^{m}, e_{\mathrm{e}}^{n}, e_{\mathrm{e}}^{r})$ represents the probability of sequentially obtaining the string of eigenvalues $\{ e_{\mathrm{e}}^{j} , e_{\mathrm{k}}^{l} , e_{\mathrm{k}}^{m}, e_{\mathrm{e}}^{n}, e_{\mathrm{e}}^{r} \}$ in an experiment in which the engine is set to run until its working substance reaches the steady state, then it is first measured after the cold isotherm, isentropically compressed, measured again, coupled to the hot bath, measured, isentropically expanded, measured, coupled to the cold bath and measured one last time. The working substance is then left alone to evolve until it reaches a steady state again and the experiment is repeated in order to obtain the probability of a different string of eigenvalues. The reason we compute $p_{w}$, $p_{q_{\mathrm{h}}}$ and $p_{q_{\mathrm{c}}}$ from the single probability distribution $p (e_{\mathrm{e}}^{j} , e_{\mathrm{k}}^{l} , e_{\mathrm{k}}^{m}, e_{\mathrm{e}}^{n}, e_{\mathrm{e}}^{r})$ is so that the statistics of the work extracted are correlated with the ones of the heat exchanged, thus making a more realistic model than if all of them were computed individually and, therefore, completely uncorrelated.

Once the probability distributions are obtained, the mean values of their respective variables follow as
\begin{equation} 
    \langle w \rangle = \int_{\mathbb{R}} dw \, w \, P (w) = - \sum_{j,l,m,n} [(e_{\mathrm{k}}^{l} - e_{\mathrm{e}}^{j}) + (e_{\mathrm{e}}^{n} - e_{\mathrm{k}}^{m})] \, p_{w} (e_{\mathrm{e}}^{j} , e_{\mathrm{k}}^{l} , e_{\mathrm{k}}^{m}, e_{\mathrm{e}}^{n}),
\end{equation}
for the work, as well as
\begin{equation}
    \langle q_{\mathrm{h}} \rangle = \int_{\mathbb{R}} d q_{\mathrm{h}} \, q_{\mathrm{h}} \, P ( q_{\mathrm{h}} ) = \sum_{l,m} (e_{\mathrm{k}}^{m} - e_{\mathrm{k}}^{l}) \, p_{q_{\mathrm{h}}} (e_{\mathrm{k}}^{l} , e_{\mathrm{k}}^{m}),
\end{equation}
and
\begin{equation}
    \langle q_{\mathrm{c}} \rangle = \int_{\mathbb{R}} d q_{\mathrm{c}} \, q_{\mathrm{c}} \, P ( q_{\mathrm{c}} ) = \sum_{n,r} (e_{\mathrm{e}}^{r} - e_{\mathrm{e}}^{n}) \, p_{q_{\mathrm{c}}} (e_{\mathrm{e}}^{n} , e_{\mathrm{e}}^{r}),
\end{equation}
for the exchanged heats. On the other hand, the corresponding variances are given by
\begin{equation}
	\sigma_{w}^{2} = \int_{\mathbb{R}} dw \, (w - \langle w \rangle)^{2} \, P (w) = \sum_{j,l,m,n} [(e_{\mathrm{k}}^{l} - e_{\mathrm{e}}^{j}) + (e_{\mathrm{e}}^{n} - e_{\mathrm{k}}^{m}) + \langle w \rangle]^{2} \, p_{w} (e_{\mathrm{e}}^{j} , e_{\mathrm{k}}^{l} , e_{\mathrm{k}}^{m}, e_{\mathrm{e}}^{n}),
\end{equation}
\begin{equation}
	\sigma_{q_{\mathrm{h}}}^{2} = \int_{\mathbb{R}} d q_{\mathrm{h}} \, (q_{\mathrm{h}} - \langle q_{\mathrm{h}} \rangle)^{2} \, P ( q_{\mathrm{h}} ) = \sum_{l,m} [ (e_{\mathrm{k}}^{m} - e_{\mathrm{k}}^{l}) - \langle q_{\mathrm{h}} \rangle]^{2} \, p_{q_{\mathrm{h}}} (e_{\mathrm{k}}^{l} , e_{\mathrm{k}}^{m}),
\end{equation}
and
\begin{equation}
	\sigma_{q_{\mathrm{c}}}^{2} = \int_{\mathbb{R}} d q_{\mathrm{c}} \, (q_{\mathrm{c}} - \langle q_{\mathrm{c}} \rangle)^{2} \, P ( q_{\mathrm{c}} ) = \sum_{n,r} [(e_{\mathrm{e}}^{r} - e_{\mathrm{e}}^{n}) - \langle q_{\mathrm{c}} \rangle]^{2} \, p_{q_{\mathrm{c}}} (e_{\mathrm{e}}^{n} , e_{\mathrm{e}}^{r}).
\end{equation}

\subsection{Two-point measurement scheme (TPM)}

To evaluate the joint distribution $p (e_{\mathrm{e}}^{j} , e_{\mathrm{k}}^{l} , e_{\mathrm{k}}^{m}, e_{\mathrm{e}}^{n}, e_{\mathrm{e}}^{r})$ with the TPM scheme \cite{mukamel_2009s,talkner_2011s}, one starts with a single copy of the working substance and makes five consecutive measurements (interspaced by the four engine cycle branches) in the eigenbasis of the corresponding Hamiltonian at that point in time. We find
\begin{equation}
    p^{\mathrm{TPM}} (e_{\mathrm{e}}^{j} , e_{\mathrm{k}}^{l} , e_{\mathrm{k}}^{m}, e_{\mathrm{e}}^{n}, e_{\mathrm{e}}^{r}) = \text{tr} \left[ \Pi_{\mathrm{e}}^{r} \, \mathcal{T}_{t_{\mathrm{c}}}^{\mathrm{c}} \left( \Pi_{\mathrm{e}}^{n} \, \mathcal{U}_{\mathrm{e}} \left\{ \Pi_{\mathrm{k}}^{m} \, \mathcal{T}_{t_{\mathrm{h}}}^{\mathrm{h}} \left[ \Pi_{\mathrm{k}}^{l} \, \mathcal{U}_{\mathrm{k}} \left( \Pi_{\mathrm{e}}^{j} \tilde{\rho}_{1} \Pi_{\mathrm{e}}^{j} \right) \Pi_{\mathrm{k}}^{l} \right] \Pi_{\mathrm{k}}^{m} \right\} \Pi_{\mathrm{e}}^{n} \right) \right].
\end{equation}
The associated marginals are therefore
\begin{equation}
    \begin{aligned}
        p_{w}^{\mathrm{TPM}} (e_{\mathrm{e}}^{j} , e_{\mathrm{k}}^{l} , e_{\mathrm{k}}^{m}, e_{\mathrm{e}}^{n}) & = \text{tr} \left( \Pi_{\mathrm{e}}^{n} \, \mathcal{U}_{\mathrm{e}} \left\{ \Pi_{\mathrm{k}}^{m} \, \mathcal{T}_{t_{\mathrm{h}}}^{\mathrm{h}} \left[ \Pi_{\mathrm{k}}^{l} \, \mathcal{U}_{\mathrm{k}} \left( \Pi_{\mathrm{e}}^{j} \tilde{\rho}_{1} \Pi_{\mathrm{e}}^{j} \right) \Pi_{\mathrm{k}}^{l} \right] \Pi_{\mathrm{k}}^{m} \right\} \right), \\
        p_{q_{\mathrm{h}}}^{\mathrm{TPM}} (e_{\mathrm{k}}^{l} , e_{\mathrm{k}}^{m}) & = \text{tr} \left\{ \Pi_{\mathrm{k}}^{m} \, \mathcal{T}_{t_{\mathrm{h}}}^{\mathrm{h}} \left[ \Pi_{\mathrm{k}}^{l} \, \mathcal{U}_{\mathrm{k}} \mathrm{D}_{\mathrm{e}} \left( \tilde{\rho}_{1} \right) \Pi_{\mathrm{k}}^{l} \right] \right\}, \\
        p_{q_{\mathrm{c}}}^{\mathrm{TPM}} (e_{\mathrm{e}}^{n} , e_{\mathrm{e}}^{r}) & = \text{tr} \left[ \Pi_{\mathrm{e}}^{r} \, \mathcal{T}_{t_{\mathrm{c}}}^{\mathrm{c}} \left( \Pi_{\mathrm{e}}^{n} \, \mathcal{U}_{\mathrm{e}} \mathrm{D}_{\mathrm{k}} \mathcal{T}_{t_{\mathrm{h}}}^{\mathrm{h}} \mathrm{D}_{\mathrm{k}} \mathcal{U}_{\mathrm{k}} \mathrm{D}_{\mathrm{e}} \left( \tilde{\rho}_{1} \right) \Pi_{\mathrm{e}}^{n} \right) \right].
    \end{aligned}
\end{equation}
The mean values for work and heat are accordingly given by 
\begin{equation}
    \begin{aligned}
        \langle w \rangle_{\mathrm{TPM}} & = \text{tr} [H_{\mathrm{k}} \, \mathcal{T}_{t_{\mathrm{h}}}^{\mathrm{h}} \mathcal{U}_{\mathrm{k}} \mathrm{D}_{\mathrm{e}} \left( \tilde{\rho}_{1} \right) ] - \text{tr} [H_{\mathrm{k}} \, \mathcal{U}_{\mathrm{k}} \mathrm{D}_{\mathrm{e}} \left( \tilde{\rho}_{1} \right) ] + \text{tr} (H_{\mathrm{e}} \tilde{\rho_{1}}) - \text{tr} [H_{\mathrm{e}} \, \mathcal{U}_{\mathrm{e}} \mathrm{D}_{\mathrm{k}} \mathcal{T}_{t_{\mathrm{h}}}^{\mathrm{h}} \mathcal{U}_{\mathrm{k}} \mathrm{D}_{\mathrm{e}} \left( \tilde{\rho}_{1} \right) ], \\
        \langle q_{{\mathrm{h}}} \rangle_{\mathrm{TPM}} & = \text{tr} [H_{\mathrm{k}} \, \mathcal{T}_{t_{\mathrm{h}}}^{\mathrm{h}} \mathcal{U}_{\mathrm{k}} \mathrm{D}_{\mathrm{e}} \left( \tilde{\rho}_{1} \right) ] - \text{tr} [H_{\mathrm{k}} \, \mathcal{U}_{\mathrm{k}} \mathrm{D}_{\mathrm{e}} \left( \tilde{\rho}_{1} \right) ], \\
        \langle q_{{\mathrm{c}}} \rangle_{\mathrm{TPM}} & = \text{tr} [H_{\mathrm{e}} \, \mathcal{T}_{t_{\mathrm{c}}}^{\mathrm{c}} \mathcal{U}_{\mathrm{e}} \mathrm{D}_{\mathrm{k}} \mathcal{T}_{t_{\mathrm{h}}}^{\mathrm{h}} \mathcal{U}_{\mathrm{k}} \mathrm{D}_{\mathrm{e}} \left( \tilde{\rho}_{1} \right) ] - \text{tr} [H_{\mathrm{e}} \, \mathcal{U}_{\mathrm{e}} \mathrm{D}_{\mathrm{k}} \mathcal{T}_{t_{\mathrm{h}}}^{\mathrm{h}} \mathcal{U}_{\mathrm{k}} \mathrm{D}_{\mathrm{e}} \left( \tilde{\rho}_{1} \right) ],
    \end{aligned}
\end{equation}
where we have used  $[\mathcal{T}_{t_{\mathrm{h,c}}}^{\mathrm{h,c}} , \mathrm{D}_{\mathrm{k,e}}] = 0$ and $ \mathrm{D}_{\mathrm{k,c}} (H_{\mathrm{k,c}}) = H_{\mathrm{k,c}}$. Interestingly, the first law  for TPM-measured engine is
\begin{equation}
    \langle w \rangle_{\mathrm{TPM}} = \langle q_{{\mathrm{h}}} \rangle_{\mathrm{TPM}} + \langle q_{{\mathrm{c}}} \rangle_{\mathrm{TPM}} + \text{tr} \left\{ H_{\mathrm{e}} \left[ \tilde{\rho_{1}} - \mathcal{T}_{t_{\mathrm{c}}}^{\mathrm{c}} \mathcal{U}_{\mathrm{e}} \mathrm{D}_{\mathrm{k}} \mathcal{T}_{t_{\mathrm{h}}}^{\mathrm{h}} \mathcal{U}_{\mathrm{k}} \mathrm{D}_{\mathrm{e}} \left( \tilde{\rho}_{1} \right) \right] \right\},
\end{equation}
where the last term measures the average deviation of the energy of the state of the working substance after a round of projective energy measurements at each point of the cycle. It is thus different from the unmeasured engine.

Moreover, the respective variances read

\begin{equation}
	\begin{aligned}
		\sigma_{w}^{2} & = \sigma_{q_{\mathrm{h}}}^{2} + \sigma_{q_{\mathrm{c}}}^{2} + \left\{ \text{tr} \left\{ H_{\mathrm{e}}^{2} \left[ \tilde{\rho_{1}} - \mathcal{T}_{t_{\mathrm{c}}}^{\mathrm{c}} \mathcal{U}_{\mathrm{e}} \mathrm{D}_{\mathrm{k}} \mathcal{T}_{t_{\mathrm{h}}}^{\mathrm{h}} \mathcal{U}_{\mathrm{k}} \mathrm{D}_{\mathrm{e}} \left( \tilde{\rho}_{1} \right) \right] \right\} - \text{tr} \left\{ H_{\mathrm{e}} \left[ \tilde{\rho_{1}} - \mathcal{T}_{t_{\mathrm{c}}}^{\mathrm{c}} \mathcal{U}_{\mathrm{e}} \mathrm{D}_{\mathrm{k}} \mathcal{T}_{t_{\mathrm{h}}}^{\mathrm{h}} \mathcal{U}_{\mathrm{k}} \mathrm{D}_{\mathrm{e}} \left( \tilde{\rho}_{1} \right) \right] \right\}^{2} \right\} \\
		& + \text{tr} [ \mathcal{T}_{t_{\mathrm{h}}}^{\mathrm{h}} \mathcal{U}_{\mathrm{k}} \mathrm{D}_{\mathrm{e}} \left( H_{\mathrm{e}}^{2} \tilde{\rho}_{1} \right) ] - \text{tr} [ H_{\mathrm{e}}^{2} \tilde{\rho}_{1} ] - 2 \langle q_{{\mathrm{h}}} \rangle \langle q_{{\mathrm{c}}} \rangle - 2 (\langle q_{{\mathrm{h}}} \rangle + \langle q_{{\mathrm{c}}} \rangle) \text{tr} \left\{ H_{\mathrm{e}} \left[ \tilde{\rho_{1}} - \mathcal{T}_{t_{\mathrm{c}}}^{\mathrm{c}} \mathcal{U}_{\mathrm{e}} \mathrm{D}_{\mathrm{k}} \mathcal{T}_{t_{\mathrm{h}}}^{\mathrm{h}} \mathcal{U}_{\mathrm{k}} \mathrm{D}_{\mathrm{e}} \left( \tilde{\rho}_{1} \right) \right] \right\} \\
		& + 2 \, \text{tr} \{ H_{\mathrm{e}} \, \mathcal{T}_{t_{\mathrm{c}}}^{\mathrm{c}} [ H_{\mathrm{e}} \mathcal{U}_{\mathrm{e}} \mathrm{D}_{\mathrm{k}} \mathcal{T}_{t_{\mathrm{h}}}^{\mathrm{h}} \mathcal{U}_{\mathrm{k}} \mathrm{D}_{\mathrm{e}} \left( \tilde{\rho}_{1} \right) ] \} + 2 \, \text{tr} \{ [H_{\mathrm{e}} \, \mathcal{U}_{\mathrm{e}} \mathrm{D}_{\mathrm{k}} \mathcal{T}_{t_{\mathrm{h}}}^{\mathrm{h}} [ H_{\mathrm{k}} \mathcal{U}_{\mathrm{k}} \mathrm{D}_{\mathrm{e}} \left( \tilde{\rho}_{1} \right) ] \} + 2 \, \text{tr} [ H_{\mathrm{k}} \, \mathcal{T}_{t_{\mathrm{h}}}^{\mathrm{h}}  \mathcal{U}_{\mathrm{k}} \mathrm{D}_{\mathrm{e}} \left( H_{\mathrm{e}} \tilde{\rho}_{1} \right) ] \\
		& - 2 \, \text{tr} [H_{\mathrm{e}} \, \mathcal{U}_{\mathrm{e}} \mathrm{D}_{\mathrm{k}} \mathcal{T}_{t_{\mathrm{h}}}^{\mathrm{h}} \mathcal{U}_{\mathrm{k}} \mathrm{D}_{\mathrm{e}} \left( H_{\mathrm{e}} \tilde{\rho}_{1} \right) ] - 2 \, \text{tr} \{ H_{\mathrm{e}} \, \mathcal{U}_{\mathrm{e}} \mathrm{D}_{\mathrm{k}} [ H_{\mathrm{e}} \mathcal{T}_{t_{\mathrm{h}}}^{\mathrm{h}} \mathcal{U}_{\mathrm{k}} \mathrm{D}_{\mathrm{e}} \left( \tilde{\rho}_{1} \right) ] \} - 2 \, \text{tr} [H_{\mathrm{k}} \, \mathcal{U}_{\mathrm{k}} \mathrm{D}_{\mathrm{e}} \left( H_{\mathrm{e}} \tilde{\rho}_{1} \right) ]
		\end{aligned}
\end{equation}
\begin{equation}
	\begin{aligned}
		\sigma_{q_{\mathrm{h}}}^{2} & = \text{tr} [H_{\mathrm{k}}^{2} \, \mathcal{T}_{t_{\mathrm{h}}}^{\mathrm{h}} \mathcal{U}_{\mathrm{k}} \mathrm{D}_{\mathrm{e}} \left( \tilde{\rho}_{1} \right) ] - \text{tr} [H_{\mathrm{k}} \, \mathcal{T}_{t_{\mathrm{h}}}^{\mathrm{h}} \mathcal{U}_{\mathrm{k}} \mathrm{D}_{\mathrm{e}} \left( \tilde{\rho}_{1} \right) ]^{2} + \text{tr} [H_{\mathrm{k}}^{2} \, \mathcal{U}_{\mathrm{k}} \mathrm{D}_{\mathrm{e}} \left( \tilde{\rho}_{1} \right) ] - \text{tr} [H_{\mathrm{k}} \, \mathcal{U}_{\mathrm{k}} \mathrm{D}_{\mathrm{e}} \left( \tilde{\rho}_{1} \right) ]^{2} \\
		& + 2 \, \text{tr} [H_{\mathrm{k}} \, \mathcal{T}_{t_{\mathrm{h}}}^{\mathrm{h}} \mathcal{U}_{\mathrm{k}} \mathrm{D}_{\mathrm{e}} \left( \tilde{\rho}_{1} \right) ] \text{tr} [H_{\mathrm{k}} \, \mathcal{U}_{\mathrm{k}} \mathrm{D}_{\mathrm{e}} \left( \tilde{\rho}_{1} \right) ] - 2 \, \text{tr} \{ H_{\mathrm{k}} \, \mathcal{T}_{t_{\mathrm{h}}}^{\mathrm{h}} [ H_{\mathrm{k}} \mathcal{U}_{\mathrm{k}} \mathrm{D}_{\mathrm{e}} \left( \tilde{\rho}_{1} \right) ] \},
	\end{aligned}
\end{equation}
\begin{equation}
	\begin{aligned}
		\sigma_{q_{\mathrm{c}}}^{2} & = \text{tr} [H_{\mathrm{e}}^{2} \, \mathcal{T}_{t_{\mathrm{c}}}^{\mathrm{c}} \mathcal{U}_{\mathrm{e}} \mathrm{D}_{\mathrm{k}} \mathcal{T}_{t_{\mathrm{h}}}^{\mathrm{h}} \mathcal{U}_{\mathrm{k}} \mathrm{D}_{\mathrm{e}} \left( \tilde{\rho}_{1} \right) ] - \text{tr} [H_{\mathrm{e}} \, \mathcal{T}_{t_{\mathrm{c}}}^{\mathrm{c}} \mathcal{U}_{\mathrm{e}} \mathrm{D}_{\mathrm{k}} \mathcal{T}_{t_{\mathrm{h}}}^{\mathrm{h}} \mathcal{U}_{\mathrm{k}} \mathrm{D}_{\mathrm{e}} \left( \tilde{\rho}_{1} \right) ]^{2} \\
		& + \text{tr} [H_{\mathrm{e}}^{2} \, \mathcal{U}_{\mathrm{e}} \mathrm{D}_{\mathrm{k}} \mathcal{T}_{t_{\mathrm{h}}}^{\mathrm{h}} \mathcal{U}_{\mathrm{k}} \mathrm{D}_{\mathrm{e}} \left( \tilde{\rho}_{1} \right) ] - \text{tr} [H_{\mathrm{e}} \, \mathcal{U}_{\mathrm{e}} \mathrm{D}_{\mathrm{k}} \mathcal{T}_{t_{\mathrm{h}}}^{\mathrm{h}} \mathcal{U}_{\mathrm{k}} \mathrm{D}_{\mathrm{e}} \left( \tilde{\rho}_{1} \right) ]^{2} \\
		& + 2 \, \text{tr} [H_{\mathrm{e}} \, \mathcal{T}_{t_{\mathrm{c}}}^{\mathrm{c}} \mathcal{U}_{\mathrm{e}} \mathrm{D}_{\mathrm{k}} \mathcal{T}_{t_{\mathrm{h}}}^{\mathrm{h}} \mathcal{U}_{\mathrm{k}} \mathrm{D}_{\mathrm{e}} \left( \tilde{\rho}_{1} \right) ] \text{tr} [H_{\mathrm{e}} \, \mathcal{U}_{\mathrm{e}} \mathrm{D}_{\mathrm{k}} \mathcal{T}_{t_{\mathrm{h}}}^{\mathrm{h}} \mathcal{U}_{\mathrm{k}} \mathrm{D}_{\mathrm{e}} \left( \tilde{\rho}_{1} \right) ] - 2 \, \text{tr} \{ H_{\mathrm{e}} \, \mathcal{T}_{t_{\mathrm{c}}}^{\mathrm{c}} [ H_{\mathrm{e}} \mathcal{U}_{\mathrm{e}} \mathrm{D}_{\mathrm{k}} \mathcal{T}_{t_{\mathrm{h}}}^{\mathrm{h}} \mathcal{U}_{\mathrm{k}} \mathrm{D}_{\mathrm{e}} \left( \tilde{\rho}_{1} \right) ] \}
	\end{aligned}
\end{equation}

\subsection{Dynamic Bayesian network approach (DBN)}

In the case of the DBN protocol \cite{lutz_2020s,par20s,mic20as,str20s,zha22s,lutz_2023s,lutz_2024s}, it is convenient to consider  five copies of the working substance \cite{lutz_2023s}: $\tilde{\rho}_{1} \otimes \tilde{\rho}_{1} \otimes \tilde{\rho}_{1} \otimes \tilde{\rho}_{1} \otimes \tilde{\rho}_{1}$ (If only $p_{w}$ and $p_{q_{\mathrm{h}}}$ are required, only four copies are necessary). Given that the quantum engine reaches a steady state after a few cycles, in order to obtain the five copies, it is enough to initialize the engine with five arbitrary states $\rho$ and wait until they all reach the steady state $\tilde{\rho}_{1}$. Let $\tilde{\rho}_{i} = \sum_{\alpha} \lambda_{i}^{\alpha} \mathrm{P}_{i}^{\alpha}$ be the spectral decomposition of the state of the working substance at corner $i$ of the engine cycle. The measurement protocol to obtain $p^{\mathrm{DBN}} (e_{\mathrm{e}}^{j} , e_{\mathrm{k}}^{l} , e_{\mathrm{k}}^{m}, e_{\mathrm{e}}^{n}, e_{\mathrm{e}}^{r})$ proceeds as follows:
\begin{enumerate}
    \item The first two copies are (locally) projectively measured in the eigenbasis of $\tilde{\rho}_{1}$ and the last four copies are left to evolve. The unnormalized state after this step is:
    \begin{equation}
        \mathrm{P}_{1}^{\alpha} \tilde{\rho}_{1} \mathrm{P}_{1}^{\alpha} \otimes \mathcal{U}_{\mathrm{k}} [ \mathrm{P}_{1}^{\alpha} \tilde{\rho}_{1} \mathrm{P}_{1}^{\alpha} / \text{tr} (\mathrm{P}_{1}^{\alpha} \tilde{\rho}_{1}) ] \otimes \tilde{\rho}_{2} \otimes \tilde{\rho}_{2} \otimes \tilde{\rho}_{2}.
    \end{equation}
    \item The second and third copies are measured in the eigenbasis of $\tilde{\rho}_{2}$, and the last three copies are left to evolve. The updated state now is:
    \begin{equation}
        \mathrm{P}_{1}^{\alpha} \tilde{\rho}_{1} \mathrm{P}_{1}^{\alpha} \otimes \mathrm{P}_{2}^{\beta} \mathcal{U}_{\mathrm{k}} [ \mathrm{P}_{1}^{\alpha} \tilde{\rho}_{1} \mathrm{P}_{1}^{\alpha} / \text{tr} (\mathrm{P}_{1}^{\alpha} \tilde{\rho}_{1}) ] \mathrm{P}_{2}^{\beta} \otimes \mathcal{T}_{t_{\mathrm{h}}}^{\mathrm{h}} [\mathrm{P}_{2}^{\beta} \tilde{\rho}_{2} \mathrm{P}_{2}^{\beta} / \text{tr} (\mathrm{P}_{2}^{\beta} \tilde{\rho}_{2}) ] \otimes \tilde{\rho}_{3} \otimes \tilde{\rho}_{3}.
    \end{equation}
    \item The third and fourth copies are measured in the eigenbasis of $\tilde{\rho}_{3}$ and the last two copies are left to evolve to obtain:
    \begin{equation}
        \mathrm{P}_{1}^{\alpha} \tilde{\rho}_{1} \mathrm{P}_{1}^{\alpha} \otimes \mathrm{P}_{2}^{\beta} \mathcal{U}_{\mathrm{k}} [ \mathrm{P}_{1}^{\alpha} \tilde{\rho}_{1} \mathrm{P}_{1}^{\alpha} / \text{tr} (\mathrm{P}_{1}^{\alpha} \tilde{\rho}_{1}) ] \mathrm{P}_{2}^{\beta} \otimes \mathrm{P}_{3}^{\gamma} \mathcal{T}_{t_{\mathrm{h}}}^{\mathrm{h}} [\mathrm{P}_{2}^{\beta} \tilde{\rho}_{2} \mathrm{P}_{2}^{\beta} / \text{tr} (\mathrm{P}_{2}^{\beta} \tilde{\rho}_{2}) ] \mathrm{P}_{3}^{\gamma} \otimes \mathcal{U}_{\mathrm{e}} [\mathrm{P}_{3}^{\gamma} \tilde{\rho}_{3} \mathrm{P}_{3}^{\gamma} / \text{tr} (\mathrm{P}_{3}^{\gamma} \tilde{\rho}_{3})] \otimes \tilde{\rho}_{4}.
    \end{equation}
    \item Copies $4$ and $5$ are measured in the eigenbasis of $\tilde{\rho}_{4}$ and the last copy is left to evolve. The state is updated to:
    \begin{equation}
        \begin{aligned}
            \mathrm{P}_{1}^{\alpha} \tilde{\rho}_{1} \mathrm{P}_{1}^{\alpha} \otimes \mathrm{P}_{2}^{\beta} \mathcal{U}_{\mathrm{k}} [ \mathrm{P}_{1}^{\alpha} \tilde{\rho}_{1} \mathrm{P}_{1}^{\alpha} / \text{tr} (\mathrm{P}_{1}^{\alpha} \tilde{\rho}_{1}) ] \mathrm{P}_{2}^{\beta} \otimes \mathrm{P}_{3}^{\gamma} \mathcal{T}_{t_{\mathrm{h}}}^{\mathrm{h}} [\mathrm{P}_{2}^{\beta} \tilde{\rho}_{2} \mathrm{P}_{2}^{\beta} / \text{tr} (\mathrm{P}_{2}^{\beta} \tilde{\rho}_{2}) ] \mathrm{P}_{3}^{\gamma} \\
            \otimes \mathrm{P}_{4}^{\delta} \mathcal{U}_{\mathrm{e}} [\mathrm{P}_{3}^{\gamma} \tilde{\rho}_{3} \mathrm{P}_{3}^{\gamma} / \text{tr} (\mathrm{P}_{3}^{\gamma} \tilde{\rho}_{3})] \mathrm{P}_{4}^{\delta} \otimes \mathcal{T}_{t_{\mathrm{c}}}^{\mathrm{c}} [ \mathrm{P}_{4}^{\delta} \tilde{\rho}_{4} \mathrm{P}_{4}^{\delta} / \text{tr} (\mathrm{P}_{4}^{\delta} \tilde{\rho}_{4}) ].
        \end{aligned}
    \end{equation}
    \item A projective measurement is performed on the last copy in the eigenbasis of $\tilde{\rho}_{1}$ to complete the cycle and get the final state: 
    \begin{equation}
        \begin{aligned}
        \mathrm{P}_{1}^{\alpha} \tilde{\rho}_{1} \mathrm{P}_{1}^{\alpha} \otimes \mathrm{P}_{2}^{\beta} \mathcal{U}_{\mathrm{k}} [ \mathrm{P}_{1}^{\alpha} \tilde{\rho}_{1} \mathrm{P}_{1}^{\alpha} / \text{tr} (\mathrm{P}_{1}^{\alpha} \tilde{\rho}_{1}) ] \mathrm{P}_{2}^{\beta} \otimes \mathrm{P}_{3}^{\gamma} \mathcal{T}_{t_{\mathrm{h}}}^{\mathrm{h}} [\mathrm{P}_{2}^{\beta} \tilde{\rho}_{2} \mathrm{P}_{2}^{\beta} / \text{tr} (\mathrm{P}_{2}^{\beta} \tilde{\rho}_{2}) ] \mathrm{P}_{3}^{\gamma} \\
        \otimes \mathrm{P}_{4}^{\delta} \mathcal{U}_{\mathrm{e}} [\mathrm{P}_{3}^{\gamma} \tilde{\rho}_{3} \mathrm{P}_{3}^{\gamma} / \text{tr} (\mathrm{P}_{3}^{\gamma} \tilde{\rho}_{3})] \mathrm{P}_{4}^{\delta} \otimes \mathrm{P}_{1}^{\epsilon} \mathcal{T}_{t_{\mathrm{c}}}^{\mathrm{c}} [\mathrm{P}_{4}^{\delta} \tilde{\rho}_{4} \mathrm{P}_{4}^{\delta} / \text{tr} (\mathrm{P}_{4}^{\delta} \tilde{\rho}_{4})] \mathrm{P}_{1}^{\epsilon}.
        \end{aligned}
    \end{equation}
\end{enumerate}
The probability of the string of measurement outcomes $\{ \lambda_{1}^{\alpha} , \lambda_{2}^{\beta}, \lambda_{3}^{\gamma}, \lambda_{4}^{\delta}, \lambda_{1}^{\epsilon} \}$ is then obtained by tracing over the un-normalized state:
\begin{equation}
    \begin{aligned}
        p (\lambda_{1}^{\alpha} , \lambda_{2}^{\beta}, \lambda_{3}^{\gamma}, \lambda_{4}^{\delta}, \lambda_{1}^{\epsilon}) & = \text{tr} (\mathrm{P}_{1}^{\alpha} \tilde{\rho}_{1}) \, \text{tr} \left\{ \mathrm{P}_{2}^{\beta} \, \mathcal{U}_{\mathrm{k}} \left[ \frac{\mathrm{P}_{1}^{\alpha} \tilde{\rho}_{1} \mathrm{P}_{1}^{\alpha}}{\text{tr} (\mathrm{P}_{1}^{\alpha} \tilde{\rho}_{1})} \right] \right\} \, \text{tr} \left\{ \mathrm{P}_{3}^{\gamma} \, \mathcal{T}_{t_{\mathrm{h}}}^{\mathrm{h}} \left[ \frac{\mathrm{P}_{2}^{\beta} \tilde{\rho}_{2} \mathrm{P}_{2}^{\beta}}{\text{tr} (\mathrm{P}_{2}^{\beta} \tilde{\rho}_{2})} \right] \right\} \\
        & \times \text{tr} \left\{ \mathrm{P}_{4}^{\delta} \, \mathcal{U}_{\mathrm{e}} \left[ \frac{\mathrm{P}_{3}^{\gamma} \tilde{\rho}_{3} \mathrm{P}_{3}^{\gamma}}{\text{tr} (\mathrm{P}_{3}^{\gamma} \tilde{\rho}_{3})} \right] \right\} \, \text{tr} \left\{ \mathrm{P}_{1}^{\epsilon} \, \mathcal{T}_{t_{\mathrm{c}}}^{\mathrm{c}} \left[ \frac{\mathrm{P}_{4}^{\delta} \tilde{\rho}_{4} \mathrm{P}_{4}^{\delta}}{\text{tr} (\mathrm{P}_{4}^{\delta} \tilde{\rho}_{4})} \right] \right\}.
    \end{aligned}
\end{equation}
Additionally, the inferred probabilities of the energy eigenvalues given a partircular measurement outcome are
\begin{equation}
    \begin{aligned}
        p (e_{\mathrm{e}}^{j} | \lambda_{1}^{\alpha}) & = \text{tr} \left[ \Pi_{\mathrm{e}}^{j} \,  \frac{\mathrm{P}_{1}^{\alpha} \tilde{\rho}_{1} \mathrm{P}_{1}^{\alpha}}{\text{tr} (\mathrm{P}_{1}^{\alpha} \tilde{\rho}_{1})} \right], \\
        p (e_{\mathrm{k}}^{l} | \lambda_{2}^{\beta}) & = \text{tr} \left[ \Pi_{\mathrm{k}}^{l} \,  \frac{\mathrm{P}_{2}^{\beta} \tilde{\rho}_{2} \mathrm{P}_{2}^{\beta}}{\text{tr} (\mathrm{P}_{2}^{\beta} \tilde{\rho}_{2})} \right], \\
        p (e_{\mathrm{k}}^{m} | \lambda_{3}^{\gamma}) & = \text{tr} \left[ \Pi_{\mathrm{k}}^{m} \,  \frac{\mathrm{P}_{3}^{\gamma} \tilde{\rho}_{3} \mathrm{P}_{3}^{\gamma}}{\text{tr} (\mathrm{P}_{3}^{\gamma} \tilde{\rho}_{3})} \right], \\
        p (e_{\mathrm{e}}^{n} | \lambda_{4}^{\delta}) & = \text{tr} \left[ \Pi_{\mathrm{e}}^{n} \,  \frac{\mathrm{P}_{4}^{\delta} \tilde{\rho}_{4} \mathrm{P}_{4}^{\delta}}{\text{tr} (\mathrm{P}_{4}^{\delta} \tilde{\rho}_{4})} \right], \\
        p (e_{\mathrm{e}}^{r} | \lambda_{1}^{\epsilon}) & = \text{tr} \left[ \Pi_{\mathrm{e}}^{r} \,  \frac{\mathrm{P}_{1}^{\epsilon} \tilde{\rho}_{1} \mathrm{P}_{1}^{\epsilon}}{\text{tr} (\mathrm{P}_{1}^{\epsilon} \tilde{\rho}_{1})} \right].
    \end{aligned}
\end{equation}
As a result, the probability distribution for the energy eigenvalues is
\begin{equation}
     p^{\mathrm{DBN}} (e_{\mathrm{e}}^{j} , e_{\mathrm{k}}^{l} , e_{\mathrm{k}}^{m}, e_{\mathrm{e}}^{n}, e_{\mathrm{e}}^{r}) = \sum_{\alpha , \beta , \gamma , \delta , \epsilon} p (\lambda_{1}^{\alpha} , \lambda_{2}^{\beta}, \lambda_{3}^{\gamma}, \lambda_{4}^{\delta}, \lambda_{1}^{\epsilon}) \, p (e_{\mathrm{e}}^{j} | \lambda_{1}^{\alpha}) \, p (e_{\mathrm{k}}^{l} | \lambda_{2}^{\beta}) \, p (e_{\mathrm{k}}^{m} | \lambda_{3}^{\gamma}) \, p (e_{\mathrm{e}}^{n} | \lambda_{4}^{\delta}) \, p (e_{\mathrm{e}}^{r} | \lambda_{1}^{\epsilon}).
\end{equation}
In this case, the marginals are
\begin{equation}
    \begin{aligned}
        p_{w}^{\mathrm{DBN}} (e_{\mathrm{e}}^{j} , e_{\mathrm{k}}^{l} , e_{\mathrm{k}}^{m}, e_{\mathrm{e}}^{n}) = & \sum_{\alpha , \beta , \gamma , \delta} \text{tr} (\mathrm{P}_{1}^{\alpha} \tilde{\rho}_{1}) \, \text{tr} \left\{ \mathrm{P}_{2}^{\beta} \, \mathcal{U}_{\mathrm{k}} \left[ \frac{\mathrm{P}_{1}^{\alpha} \tilde{\rho}_{1} \mathrm{P}_{1}^{\alpha}}{\text{tr} (\mathrm{P}_{1}^{\alpha} \tilde{\rho}_{1})} \right] \right\} \, \text{tr} \left\{ \mathrm{P}_{3}^{\gamma} \, \mathcal{T}_{t_{\mathrm{h}}}^{\mathrm{h}} \left[ \frac{\mathrm{P}_{2}^{\beta} \tilde{\rho}_{2} \mathrm{P}_{2}^{\beta}}{\text{tr} (\mathrm{P}_{2}^{\beta} \tilde{\rho}_{2})} \right] \right\} \, \text{tr} \left\{ \mathrm{P}_{4}^{\delta} \, \mathcal{U}_{\mathrm{e}} \left[ \frac{\mathrm{P}_{3}^{\gamma} \tilde{\rho}_{3} \mathrm{P}_{3}^{\gamma}}{\text{tr} (\mathrm{P}_{3}^{\gamma} \tilde{\rho}_{3})} \right] \right\}, \\
        & \times \text{tr} \left[ \Pi_{\mathrm{e}}^{j} \,  \frac{\mathrm{P}_{1}^{\alpha} \tilde{\rho}_{1} \mathrm{P}_{1}^{\alpha}}{\text{tr} (\mathrm{P}_{1}^{\alpha} \tilde{\rho}_{1})} \right] \, \text{tr} \left[ \Pi_{\mathrm{k}}^{l} \, \frac{\mathrm{P}_{2}^{\beta} \tilde{\rho}_{2} \mathrm{P}_{2}^{\beta}}{\text{tr} (\mathrm{P}_{2}^{\beta} \tilde{\rho}_{2})} \right] \, \text{tr} \left[ \Pi_{\mathrm{k}}^{m} \,  \frac{\mathrm{P}_{3}^{\gamma} \tilde{\rho}_{3} \mathrm{P}_{3}^{\gamma}}{\text{tr} (\mathrm{P}_{3}^{\gamma} \tilde{\rho}_{3})} \right] \, \text{tr} \left[ \Pi_{\mathrm{e}}^{n} \,  \frac{\mathrm{P}_{4}^{\delta} \tilde{\rho}_{4} \mathrm{P}_{4}^{\delta}}{\text{tr} (\mathrm{P}_{4}^{\delta} \tilde{\rho}_{4})} \right], \\
        p_{q_{\mathrm{h}}}^{\mathrm{DBN}} (e_{\mathrm{k}}^{l} , e_{\mathrm{k}}^{m}) = & \sum_{\beta , \gamma} \text{tr} \left[ \mathrm{P}_{3}^{\gamma} \, \mathcal{T}_{t_{\mathrm{h}}}^{\mathrm{h}} \left( \mathrm{P}_{2}^{\beta} \tilde{\rho}_{2} \mathrm{P}_{2}^{\beta} \right) \right] \, \text{tr} \left[ \Pi_{\mathrm{k}}^{l} \,  \frac{\mathrm{P}_{2}^{\beta} \tilde{\rho}_{2} \mathrm{P}_{2}^{\beta}}{\text{tr} (\mathrm{P}_{2}^{\beta} \tilde{\rho}_{2})} \right] \, \text{tr} \left[ \Pi_{\mathrm{k}}^{m} \,  \frac{\mathrm{P}_{3}^{\gamma} \tilde{\rho}_{3} \mathrm{P}_{3}^{\gamma}}{\text{tr} (\mathrm{P}_{3}^{\gamma} \tilde{\rho}_{3})} \right], \\
        p_{q_{\mathrm{c}}}^{\mathrm{DBN}} (e_{\mathrm{e}}^{n} , e_{\mathrm{e}}^{r}) = & \sum_{\delta , \epsilon} \text{tr} \left[ \mathrm{P}_{1}^{\epsilon} \, \mathcal{T}_{t_{\mathrm{c}}}^{\mathrm{c}} \left( \mathrm{P}_{4}^{\delta} \tilde{\rho}_{4} \mathrm{P}_{4}^{\delta} \right) \right] \, \text{tr} \left[ \Pi_{\mathrm{e}}^{n} \,  \frac{\mathrm{P}_{4}^{\delta} \tilde{\rho}_{4} \mathrm{P}_{4}^{\delta}}{\text{tr} (\mathrm{P}_{4}^{\delta} \tilde{\rho}_{4})} \right] \, \text{tr} \left[ \Pi_{\mathrm{e}}^{r} \,  \frac{\mathrm{P}_{1}^{\epsilon} \tilde{\rho}_{1} \mathrm{P}_{1}^{\epsilon}}{\text{tr} (\mathrm{P}_{1}^{\epsilon} \tilde{\rho}_{1})} \right].
    \end{aligned}
\end{equation}
The mean values of work and heat follow as
\begin{equation}
    \begin{aligned}
        \langle w \rangle_{{\mathrm{DBN}}} & = \text{tr} (H_{\mathrm{k}} \tilde{\rho}_{3}) - \text{tr} (H_{\mathrm{k}} \tilde{\rho}_{2}) + \text{tr} (H_{\mathrm{e}} \tilde{\rho}_{1}) - \text{tr} (H_{\mathrm{e}} \tilde{\rho}_{4}) = W \\
        \langle q_{\mathrm{h}} \rangle_{{\mathrm{DBN}}} & = \text{tr} (H_{\mathrm{k}} \tilde{\rho}_{3}) - \text{tr} (H_{\mathrm{k}} \tilde{\rho}_{2}) = Q_{\mathrm{h}} \\
        \langle q_{\mathrm{c}} \rangle_{{\mathrm{DBN}}} & = \text{tr} (H_{\mathrm{e}} \tilde{\rho}_{1}) - \text{tr} (H_{\mathrm{e}} \tilde{\rho}_{4}) = Q_{\mathrm{c}}.
    \end{aligned}
\end{equation}
The variances furthermore read
\begin{equation}
	\begin{aligned}
		\sigma_{w}^{2} & = \sigma_{q_{\mathrm{h}}}^{2} + \sigma_{q_{\mathrm{c}}}^{2} - 2 \langle q_{\mathrm{h}} \rangle \langle q_{\mathrm{c}} \rangle + 2 \sum_{\delta , \epsilon} \text{tr} \left[ \mathrm{P}_{1}^{\epsilon} \, \mathcal{T}_{t_{\mathrm{c}}}^{\mathrm{c}} \left( \mathrm{P}_{4}^{\delta} \tilde{\rho}_{4} \mathrm{P}_{4}^{\delta} \right) \right] \, \text{tr} \left[ H_{\mathrm{e}} \,  \frac{\mathrm{P}_{4}^{\delta} \tilde{\rho}_{4} \mathrm{P}_{4}^{\delta}}{\text{tr} (\mathrm{P}_{4}^{\delta} \tilde{\rho}_{4})} \right] \, \text{tr} \left[ H_{\mathrm{e}} \,  \frac{\mathrm{P}_{1}^{\epsilon} \tilde{\rho}_{1} \mathrm{P}_{1}^{\epsilon}}{\text{tr} (\mathrm{P}_{1}^{\epsilon} \tilde{\rho}_{1})} \right] \\
		& - 2 \sum_{\alpha , \beta , \gamma , \delta} \text{tr} (\mathrm{P}_{1}^{\alpha} \tilde{\rho}_{1}) \, \text{tr} \left\{ \mathrm{P}_{2}^{\beta} \, \mathcal{U}_{\mathrm{k}} \left[ \frac{\mathrm{P}_{1}^{\alpha} \tilde{\rho}_{1} \mathrm{P}_{1}^{\alpha}}{\text{tr} (\mathrm{P}_{1}^{\alpha} \tilde{\rho}_{1})} \right] \right\} \, \text{tr} \left\{ \mathrm{P}_{3}^{\gamma} \, \mathcal{T}_{t_{\mathrm{h}}}^{\mathrm{h}} \left[ \frac{\mathrm{P}_{2}^{\beta} \tilde{\rho}_{2} \mathrm{P}_{2}^{\beta}}{\text{tr} (\mathrm{P}_{2}^{\beta} \tilde{\rho}_{2})} \right] \right\} \, \text{tr} \left\{ \mathrm{P}_{4}^{\delta} \, \mathcal{U}_{\mathrm{e}} \left[ \frac{\mathrm{P}_{3}^{\gamma} \tilde{\rho}_{3} \mathrm{P}_{3}^{\gamma}}{\text{tr} (\mathrm{P}_{3}^{\gamma} \tilde{\rho}_{3})} \right] \right\} \\
		& \times \text{tr} \left[ H_{\mathrm{e}} \,  \frac{\mathrm{P}_{1}^{\alpha} \tilde{\rho}_{1} \mathrm{P}_{1}^{\alpha}}{\text{tr} (\mathrm{P}_{1}^{\alpha} \tilde{\rho}_{1})} \right] \, \text{tr} \left[ H_{\mathrm{e}} \,  \frac{\mathrm{P}_{4}^{\delta} \tilde{\rho}_{4} \mathrm{P}_{4}^{\delta}}{\text{tr} (\mathrm{P}_{4}^{\delta} \tilde{\rho}_{4})} \right] \\
		& + 2 \sum_{\beta , \gamma , \delta} \text{tr} \left[ \mathrm{P}_{3}^{\gamma} \, \mathcal{T}_{t_{\mathrm{h}}}^{\mathrm{h}} \left( \mathrm{P}_{2}^{\beta} \tilde{\rho}_{2} \mathrm{P}_{2}^{\beta} \right) \right] \, \text{tr} \left\{ \mathrm{P}_{4}^{\delta} \, \mathcal{U}_{\mathrm{e}} \left[ \frac{\mathrm{P}_{3}^{\gamma} \tilde{\rho}_{3} \mathrm{P}_{3}^{\gamma}}{\text{tr} (\mathrm{P}_{3}^{\gamma} \tilde{\rho}_{3})} \right] \right\} \, \text{tr} \left[ H_{\mathrm{k}} \, \frac{\mathrm{P}_{2}^{\beta} \tilde{\rho}_{2} \mathrm{P}_{2}^{\beta}}{\text{tr} (\mathrm{P}_{2}^{\beta} \tilde{\rho}_{2})} \right] \, \text{tr} \left[ H_{\mathrm{e}} \,\frac{\mathrm{P}_{4}^{\delta} \tilde{\rho}_{4} \mathrm{P}_{4}^{\delta}}{\text{tr} (\mathrm{P}_{4}^{\delta} \tilde{\rho}_{4})} \right] \\
		& + 2 \sum_{\alpha , \beta , \gamma} \text{tr} (\mathrm{P}_{1}^{\alpha} \tilde{\rho}_{1}) \, \text{tr} \left\{ \mathrm{P}_{2}^{\beta} \, \mathcal{U}_{\mathrm{k}} \left[ \frac{\mathrm{P}_{1}^{\alpha} \tilde{\rho}_{1} \mathrm{P}_{1}^{\alpha}}{\text{tr} (\mathrm{P}_{1}^{\alpha} \tilde{\rho}_{1})} \right] \right\} \, \text{tr} \left\{ \mathrm{P}_{3}^{\gamma} \, \mathcal{T}_{t_{\mathrm{h}}}^{\mathrm{h}} \left[ \frac{\mathrm{P}_{2}^{\beta} \tilde{\rho}_{2} \mathrm{P}_{2}^{\beta}}{\text{tr} (\mathrm{P}_{2}^{\beta} \tilde{\rho}_{2})} \right] \right\} \, \text{tr} \left[ H_{\mathrm{e}} \,  \frac{\mathrm{P}_{1}^{\alpha} \tilde{\rho}_{1} \mathrm{P}_{1}^{\alpha}}{\text{tr} (\mathrm{P}_{1}^{\alpha} \tilde{\rho}_{1})} \right] \, \text{tr} \left[ H_{\mathrm{k}} \,  \frac{\mathrm{P}_{3}^{\gamma} \tilde{\rho}_{3} \mathrm{P}_{3}^{\gamma}}{\text{tr} (\mathrm{P}_{3}^{\gamma} \tilde{\rho}_{3})} \right] \\
		& - 2 \sum_{\alpha , \beta} \text{tr} (\mathrm{P}_{1}^{\alpha} \tilde{\rho}_{1}) \, \text{tr} \left\{ \mathrm{P}_{2}^{\beta} \, \mathcal{U}_{\mathrm{k}} \left[ \frac{\mathrm{P}_{1}^{\alpha} \tilde{\rho}_{1} \mathrm{P}_{1}^{\alpha}}{\text{tr} (\mathrm{P}_{1}^{\alpha} \tilde{\rho}_{1})} \right] \right\} \, \text{tr} \left[ H_{\mathrm{e}} \,  \frac{\mathrm{P}_{1}^{\alpha} \tilde{\rho}_{1} \mathrm{P}_{1}^{\alpha}}{\text{tr} (\mathrm{P}_{1}^{\alpha} \tilde{\rho}_{1})} \right] \, \text{tr} \left[ H_{\mathrm{k}} \, \frac{\mathrm{P}_{2}^{\beta} \tilde{\rho}_{2} \mathrm{P}_{2}^{\beta}}{\text{tr} (\mathrm{P}_{2}^{\beta} \tilde{\rho}_{2})} \right] \\
		& - 2 \sum_{\gamma , \delta} \text{tr} \left[ \mathrm{P}_{4}^{\delta} \, \mathcal{U}_{\mathrm{e}} \left( \mathrm{P}_{3}^{\gamma} \tilde{\rho}_{3} \mathrm{P}_{3}^{\gamma} \right) \right] \, \text{tr} \left[ H_{\mathrm{k}} \,  \frac{\mathrm{P}_{3}^{\gamma} \tilde{\rho}_{3} \mathrm{P}_{3}^{\gamma}}{\text{tr} (\mathrm{P}_{3}^{\gamma} \tilde{\rho}_{3})} \right] \, \text{tr} \left[ H_{\mathrm{e}} \,\frac{\mathrm{P}_{4}^{\delta} \tilde{\rho}_{4} \mathrm{P}_{4}^{\delta}}{\text{tr} (\mathrm{P}_{4}^{\delta} \tilde{\rho}_{4})} \right].
	\end{aligned}
\end{equation}

\begin{equation}
	\begin{aligned}
		\sigma_{q_{\mathrm{h}}}^{2} & = \text{tr} (H_{\mathrm{k}}^{2} \tilde{\rho}_{3}) - \text{tr} (H_{\mathrm{k}} \tilde{\rho}_{3})^{2} + \text{tr} (H_{\mathrm{k}}^{2} \tilde{\rho}_{2}) - \text{tr} (H_{\mathrm{k}} \tilde{\rho}_{2})^{2} \\
		& + 2 \, \text{tr} (H_{\mathrm{k}} \tilde{\rho}_{3}) \text{tr} (H_{\mathrm{k}} \tilde{\rho}_{2})  - 2 \sum_{\beta , \gamma} \text{tr} \left[ \mathrm{P}_{3}^{\gamma} \, \mathcal{T}_{t_{\mathrm{h}}}^{\mathrm{h}} \left( \mathrm{P}_{2}^{\beta} \tilde{\rho}_{2} \mathrm{P}_{2}^{\beta} \right) \right] \, \text{tr} \left[ H_{\mathrm{k}} \,  \frac{\mathrm{P}_{2}^{\beta} \tilde{\rho}_{2} \mathrm{P}_{2}^{\beta}}{\text{tr} (\mathrm{P}_{2}^{\beta} \tilde{\rho}_{2})} \right] \, \text{tr} \left[ H_{\mathrm{k}} \,  \frac{\mathrm{P}_{3}^{\gamma} \tilde{\rho}_{3} \mathrm{P}_{3}^{\gamma}}{\text{tr} (\mathrm{P}_{3}^{\gamma} \tilde{\rho}_{3})} \right].
	\end{aligned}
\end{equation}
\begin{equation}
	\begin{aligned}
		\sigma_{q_{\mathrm{c}}}^{2} & = \text{tr} (H_{\mathrm{e}}^{2} \tilde{\rho}_{1}) - \text{tr} (H_{\mathrm{e}} \tilde{\rho}_{1})^{2} + \text{tr} (H_{\mathrm{e}}^{2} \tilde{\rho}_{4}) - \text{tr} (H_{\mathrm{e}} \tilde{\rho}_{4})^{2} \\
		& + 2 \, \text{tr} (H_{\mathrm{e}} \tilde{\rho}_{1}) \text{tr} (H_{\mathrm{e}} \tilde{\rho}_{4}) - 2 \sum_{\delta , \epsilon} \text{tr} \left[ \mathrm{P}_{1}^{\epsilon} \, \mathcal{T}_{t_{\mathrm{c}}}^{\mathrm{c}} \left( \mathrm{P}_{4}^{\delta} \tilde{\rho}_{4} \mathrm{P}_{4}^{\delta} \right) \right] \, \text{tr} \left[ H_{\mathrm{e}} \,  \frac{\mathrm{P}_{4}^{\delta} \tilde{\rho}_{4} \mathrm{P}_{4}^{\delta}}{\text{tr} (\mathrm{P}_{4}^{\delta} \tilde{\rho}_{4})} \right] \, \text{tr} \left[ H_{\mathrm{e}} \,  \frac{\mathrm{P}_{1}^{\epsilon} \tilde{\rho}_{1} \mathrm{P}_{1}^{\epsilon}}{\text{tr} (\mathrm{P}_{1}^{\epsilon} \tilde{\rho}_{1})} \right].
	\end{aligned}
\end{equation}

\section{Equality between the TPM and DBN probability distributions} 

In this section, we will show that the conditions
\begin{equation}
    \begin{cases}
        & \mathcal{U}_{\mathrm{k}} \mathrm{D}_{\mathrm{e}} (\rho) = \mathrm{D}_{\mathrm{k}} \mathcal{U}_{\mathrm{k}} \mathrm{D}_{\mathrm{e}} (\rho) \quad \text{or} \quad t_{\mathrm{h}} \to \infty \\
        & \mathcal{U}_{\mathrm{e}} \mathrm{D}_{\mathrm{k}} (\rho) = \mathrm{D}_{\mathrm{e}} \mathcal{U}_{\mathrm{e}} \mathrm{D}_{\mathrm{k}} (\rho) \quad \text{or} \quad t_{\mathrm{c}} \to \infty.
    \end{cases}
    \label{eqsm:conditionTPM}
\end{equation}
imply that the state of the working substance is diagonal in the basis of the corresponding Hamiltonian at each corner of the cycle and, therefore, that the probability distributions computed with the TPM  or DBN schemes are the same. In order to do so, first note that generically the generalized amplitude damping map (as a function of the thermalization rate instead of the time) can be written as \cite{srikanth_2024}
\begin{equation}
    \mathcal{T}_{\lambda} (\rho) = (1 - \lambda) \rho + \lambda \, \sigma - 2 \sqrt{1 - \lambda} (1 - \sqrt{1-\lambda}) \sum_{j = 1}^{d} p_{j} \mathcal{D}_{\Pi^{j}} (\rho),
\end{equation}
where $\lambda \in [0,1]$, $\sigma  =\sum_{j} p_{j} \Pi^{j}$ is the unique fixed point of $\mathcal{T}_{\lambda}$, and $\mathcal{D}_{\Pi^{j}} (\cdot) = \Pi^{j} \cdot \Pi^{j} - \{ \Pi^{j} , \cdot \} / 2$ is the usual Lindblad dissipator. In the case of the map describing the interaction with the cold reservoir, $\mathcal{T}_{\lambda}^{\mathrm{c}}$, the fixed point is $\sigma = \text{exp} [- H_{\mathrm{e}} (t_{\mathrm{e}}) / T_{\mathrm{c}}] / \text{tr} \{ \text{exp} [- H_{\mathrm{e}} (t_{\mathrm{e}}) / T_{\mathrm{c}}] \}$, and the projectors are $\{ \Pi_{\mathrm{e}}^{j} \}_{j}$. Analogously, for the interaction with the hot reservoir, $\mathcal{T}_{\lambda}^{\mathrm{h}}$, $\sigma = \text{exp} [- H_{\mathrm{k}} (t_{\mathrm{k}}) / T_{\mathrm{h}}] / \text{tr} \{ \text{exp} [- H_{\mathrm{k}} (t_{\mathrm{k}}) / T_{\mathrm{h}}] \}$ and the projectors are $\{ \Pi_{\mathrm{k}}^{j} \}_{j}$. It is straightforward to check that, if $\mathrm{D} (\cdot) =\sum_{j} \Pi^{j} \cdot \Pi^{j}$ is the dephasing map in the eigenbasis of the fixed point of $\mathcal{T}_{\lambda}$, then $[\mathrm{D} , \mathcal{T}_{\lambda}] = 0$. Thus, diagonal states in that basis are preserved by $\mathcal{T}_{\lambda}$. On the other hand, if $\rho$ was not diagonal in that basis, then $\mathcal{T}_{\lambda} (\rho)$ is also not diagonal unless $\lambda = 1$. Therfore, $\mathcal{T}_{\lambda} (\rho)$ is diagonal if $\rho$ was already diagonal or if $\lambda = 1$, as it was stated in the main text. Now, assuming the conditions in Eq.~\eqref{eqsm:conditionTPM} hold, a direct computation shows that diagonal states in the basis of $H_{\mathrm{e}} (t_{\mathrm{e}})$ are preserved during the engine cycle, $\Lambda_{1} D_{\mathrm{e}} (\rho) = D_{\mathrm{e}} \Lambda_{1} D_{\mathrm{e}} (\rho)$, which immediately leads to $\Lambda_{1}^{n} D_{\mathrm{e}} (\rho) = D_{\mathrm{e}} \Lambda_{1}^{n} D_{\mathrm{e}} (\rho)$. By taking the limit $n \to \infty$ and using the fact that the fixed point of $\Lambda_{1}$ is unique, the result is that the steady state $\tilde{\rho}_{1}$ is diagonal in the eigenbasis of $H_{\mathrm{e}} (t_{\mathrm{e}})$, $ \tilde{\rho}_{1} = D_{\mathrm{e}} (\tilde{\rho}_{1})$. By the first line of Eq.~\eqref{eqsm:conditionTPM}, the state of the working substance is diagonal in the eigenbasis of $H_{\mathrm{k}} (t_{\mathrm{k}})$ after the isentropic compression and, therefore, after the hot isothermal isochore too. The second line in Eq.~\eqref{eqsm:conditionTPM} then implies that it is diagonal in the eigenbasis of $H_{\mathrm{e}} (t_{\mathrm{e}})$ after the isentropic expansion. Consequently, the state of the working substance is diagonal in the corresponding Hamiltonian basis in the four corners of the cycle and, as a result, the TPM  and DBN protocols are equivalent.

\begin{figure}[!htb]{%
        \includegraphics[width=0.5\linewidth]{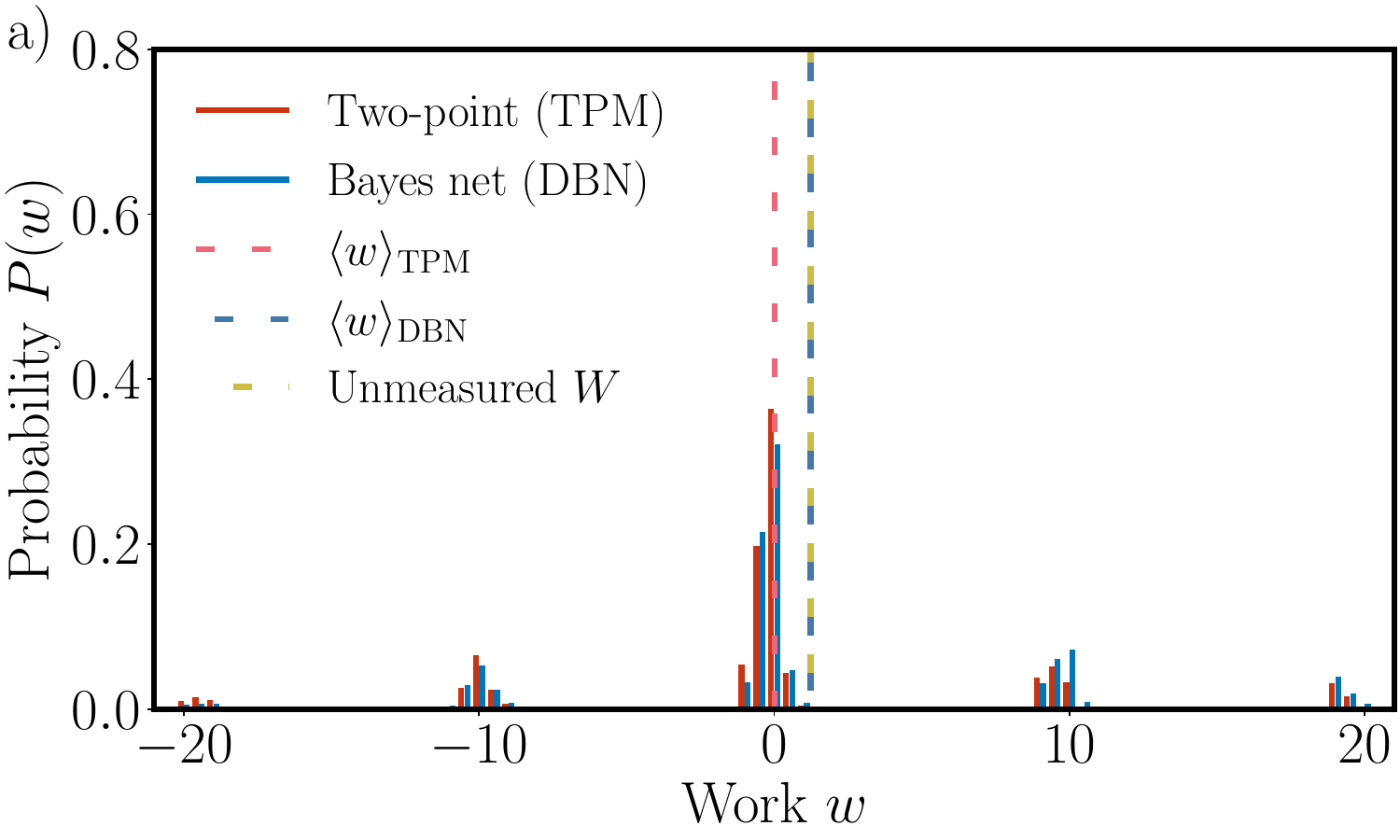}%
    } \\ 
    \label{figsm:corb}{%
        \includegraphics[width=0.48\linewidth]{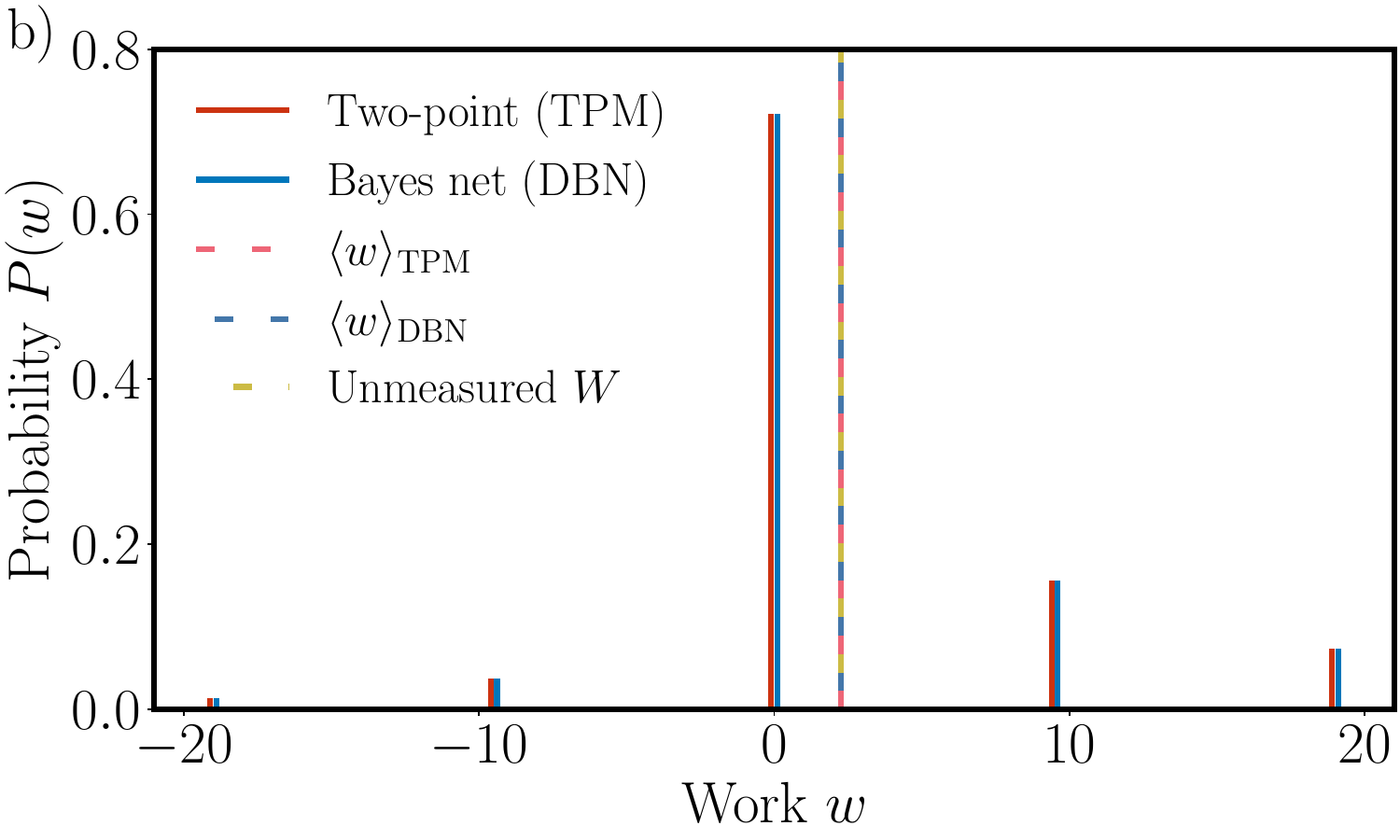}%
    } \hspace{0.02\linewidth}
    \label{figsm:corc}{%
        \includegraphics[width=0.48\linewidth]{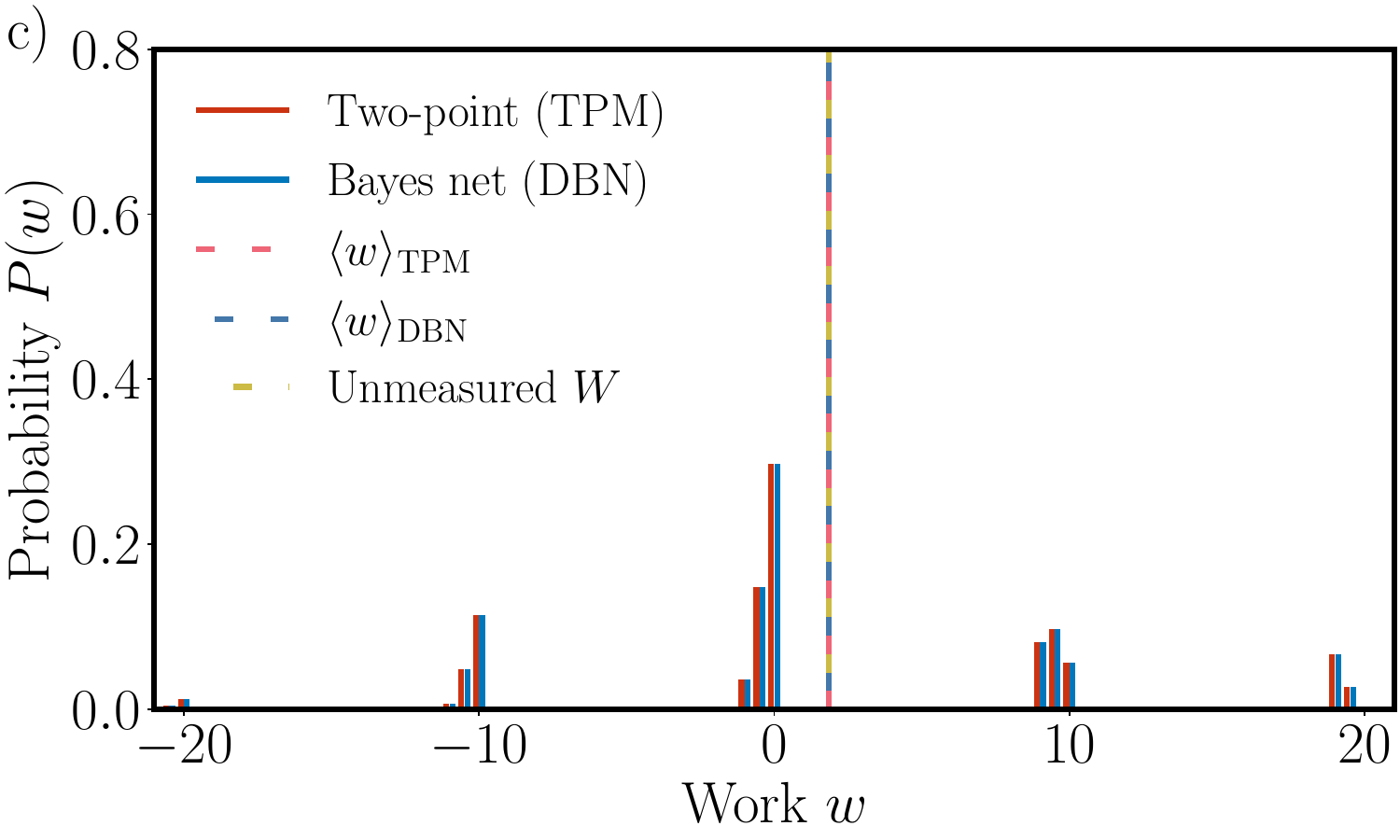}%
    } 
    
	\caption{Work probability distributions $P (w)$ of the engine cycle computed using TPM (red) and DBN (blue), with their respective means, $\langle w \rangle_{\mathrm{TPM}}$ and $\langle w \rangle_{\mathrm{DBN}}$, compared to the unmeasured value of work $W$. a) In the generic case, where coherences in the energy basis are dynamically induced, $g = 9$, and the working substance does not fully thermalize, $t_{\mathrm{h}} = t_{\mathrm{c}} = 0.92$, both probability distributions are different and only the one computed with DBN reproduces the correct mean value for the work. b) In the absence of dynamically induced coherences, $g = 0$ and $t_{\mathrm{h}} = t_{\mathrm{c}} = 0.92$, or c) if the working substance thermalizes, $g = 9$ and $t_{\mathrm{h} , \mathrm{c}} \to \infty$, TPM and DBN are equivalent. Parameters are $\omega_{\mathrm{h}} = 10$, $T_{\mathrm{h}} = 14$, $\omega_{\mathrm{c}} = 0.5$, $T_{\mathrm{c}} = 0.1$, $g = 9$, $t_{\mathrm{h}} = t_{\mathrm{c}} = 0.92$.}
    \label{figsm:pw}
\end{figure}

In Fig.~\ref{figsm:pw} we show three generic examples of the work probability distributions $P_{\text{TPM}}$ and $P_{\text{DBN}}$. Note they coincide if $g = 0$ or if $t_{\mathrm{h} , \mathrm{c}} \to \infty$.

\section{Comparison between the TPM and DBN protocols}

\subsection{Kullback-Leibler divergence and relative entropy of coherence}
\begin{figure}[!htb]
    {%
        \includegraphics[width=0.5\linewidth]{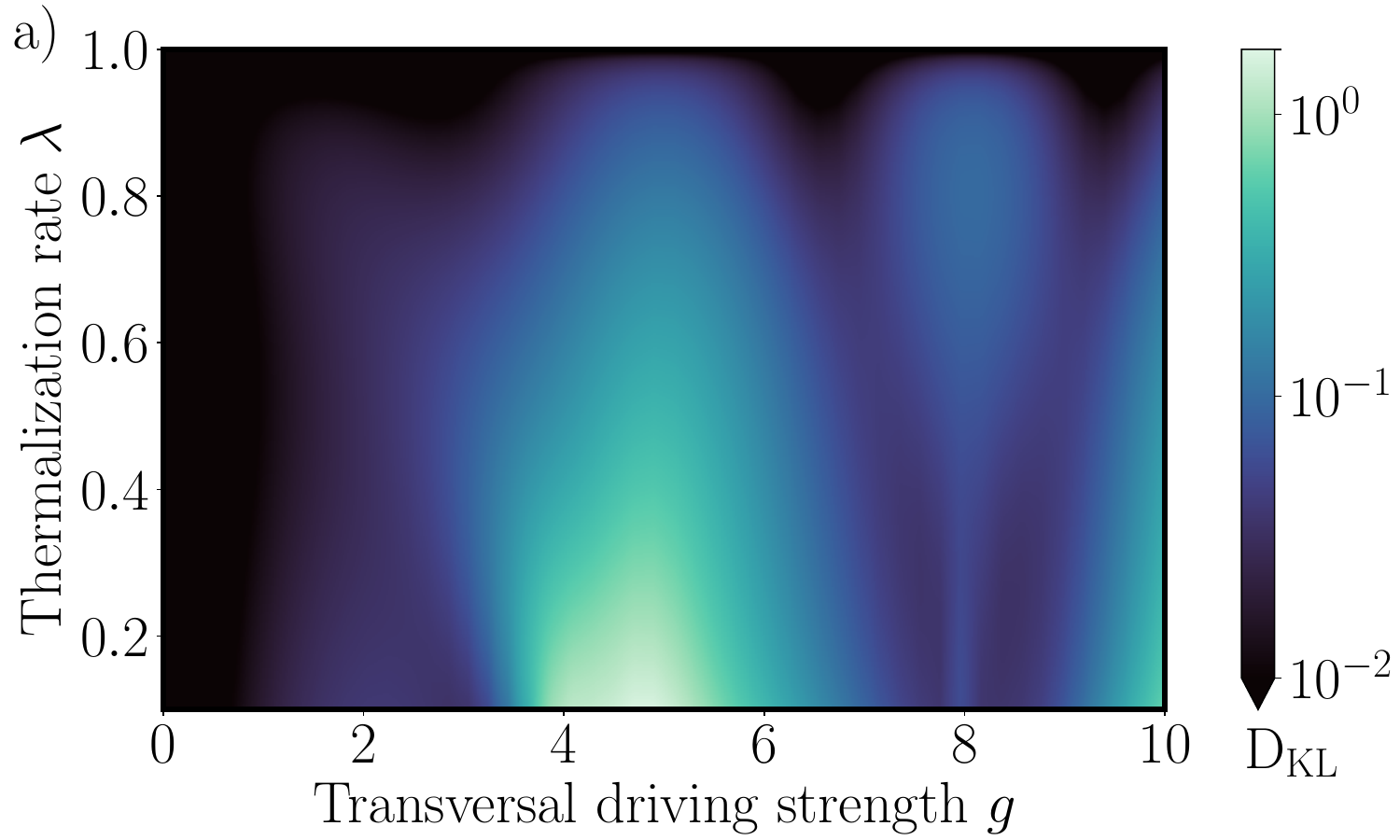}%
    } \\ 
    \label{figsm:corb}{%
        \includegraphics[width=0.48\linewidth]{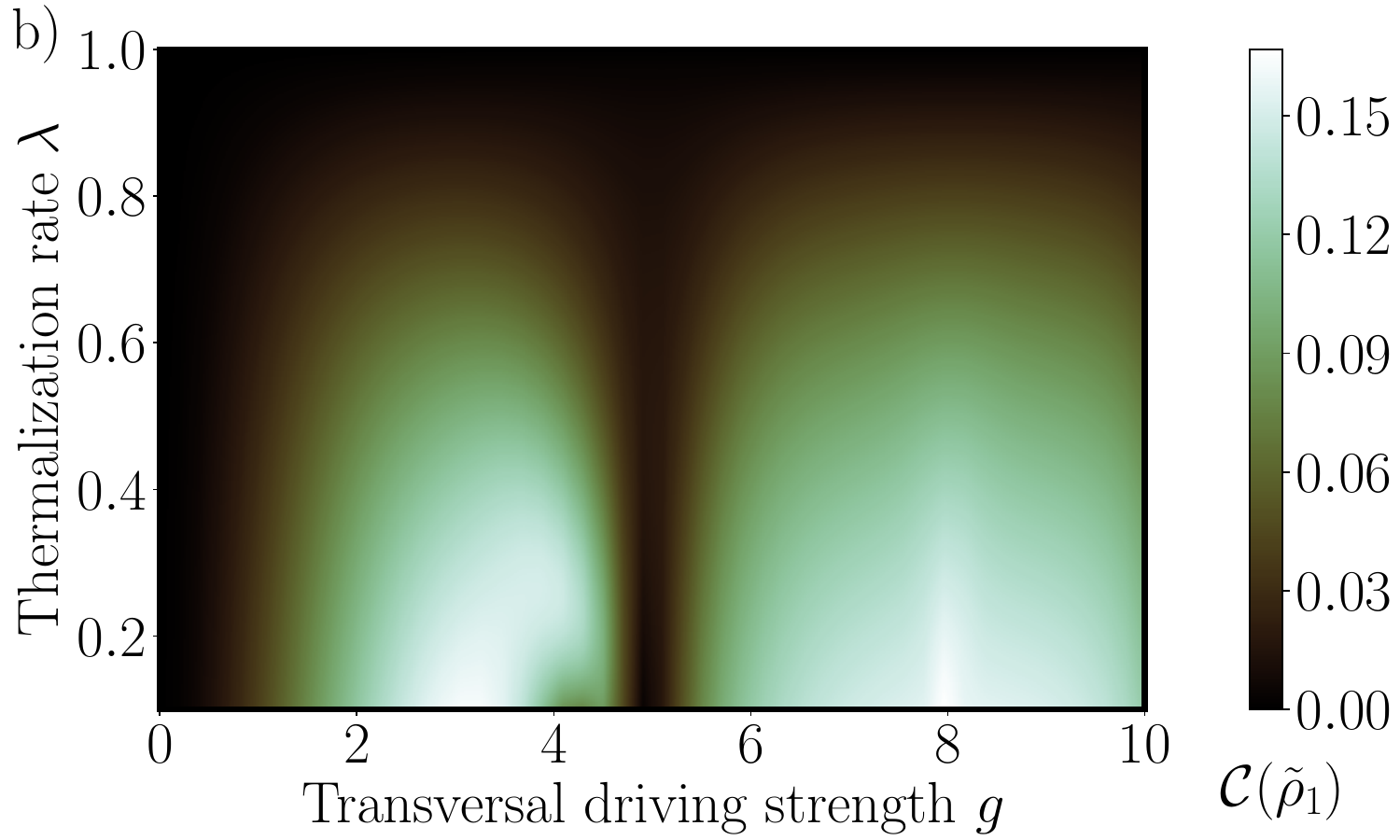}%
    } \hspace{0.02\linewidth}
    \label{figsm:corc}{%
        \includegraphics[width=0.48\linewidth]{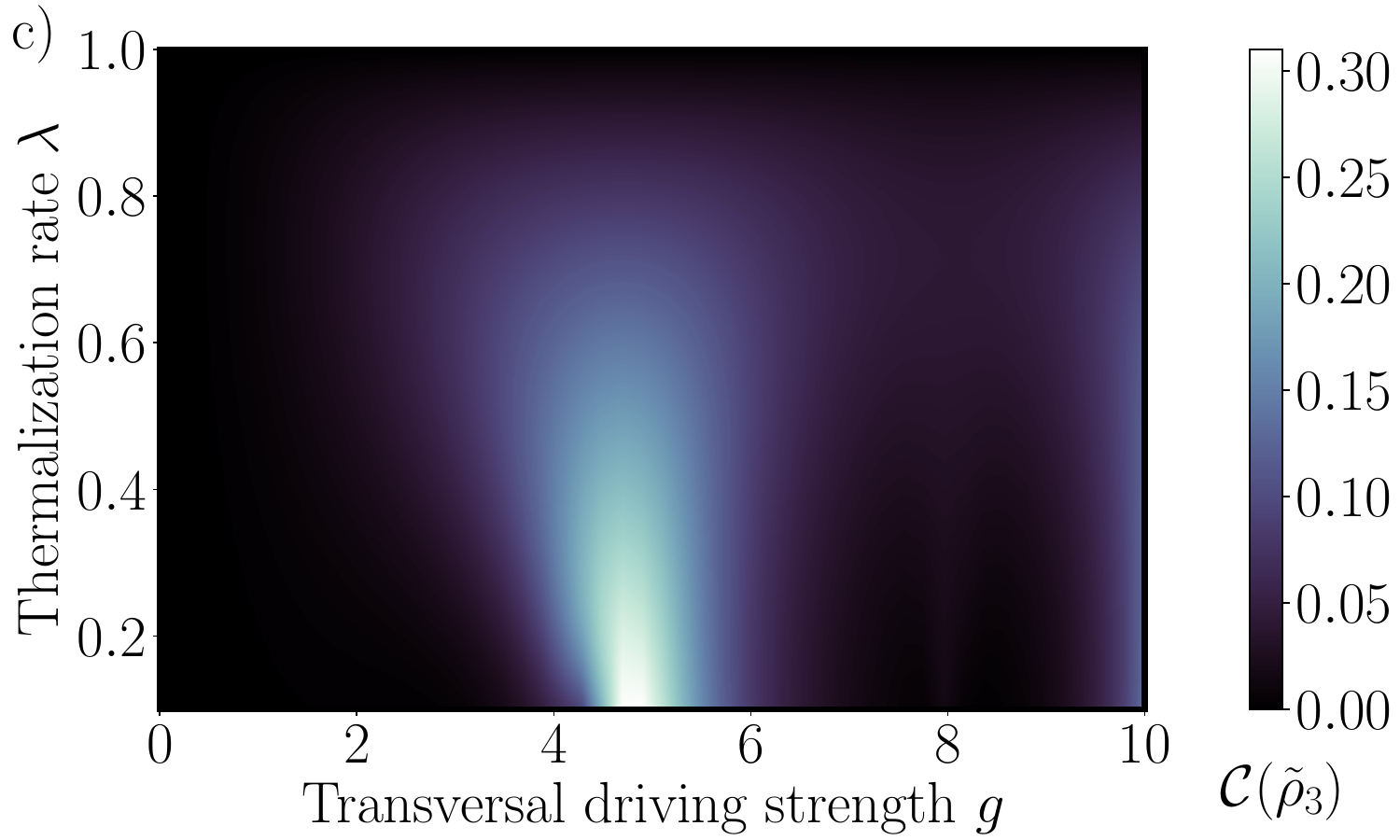}%
    } 
    \caption{a) Kullback-Leibler divergence, $D_\textrm{KL} (P_{\textrm{DBN}} || P_{\textrm{TPM}})$, between the DBN and the TPM work probability distributions. b) Relative entropy of coherence $\mathcal{C}$ of the steady state before the isentropic compression step, $\tilde{\rho}_{1}$. c)  Relative entropy of coherence $\mathcal{C}$ of the steady state before the isentropic expansion step, $\tilde{\rho}_{3}$. The regions in which the two probability distributions differ the most (a) coincide with the zones where the steady state has the most coherence (bc). Parameters  are $\omega_{\mathrm{h}} = 10$, $T_{\mathrm{h}} = 14$, $\omega_{\mathrm{c}} = 0.5$, $T_{\mathrm{c}} = 0.1$, and $\lambda_{\mathrm{c}} = \lambda_{\mathrm{h}} = \lambda$.}\label{figsm:corSM}
\end{figure}

In the main text, we compared the entropic distance (or Kullback-Leibler divergence), $D_\textrm{KL} (P_{\textrm{DBN}} || P_{\textrm{TPM}})=\int dw\, P_{\textrm{DBN}}(w) \ln[P_{\textrm{DBN}}(w)/P_{\textrm{TPM}}(w)]$  and  the relative entropy of coherence, $\mathcal{C}(\tilde{\rho})=\text{tr}\{\text{D}(\tilde{\rho}) \{ \ln[\text{D}(\tilde{\rho})]-\ln(\tilde{\rho})\} \}$ for a three-level quantum Otto cycle, as a function of the thermalization rate $\lambda$, for a fixed value of the transverse driving strength $g$ (Fig.~1). In Fig.~\ref{figsm:corSM}, we complement these results by showing contour plots for the same quantities when the transverse driving strength $g$ is varied.

\clearpage

\subsection{Backaction and trace distance between measured and unmeasured steady states}
\begin{figure}[!htb]
    \centering
    \includegraphics[width=0.7\linewidth]{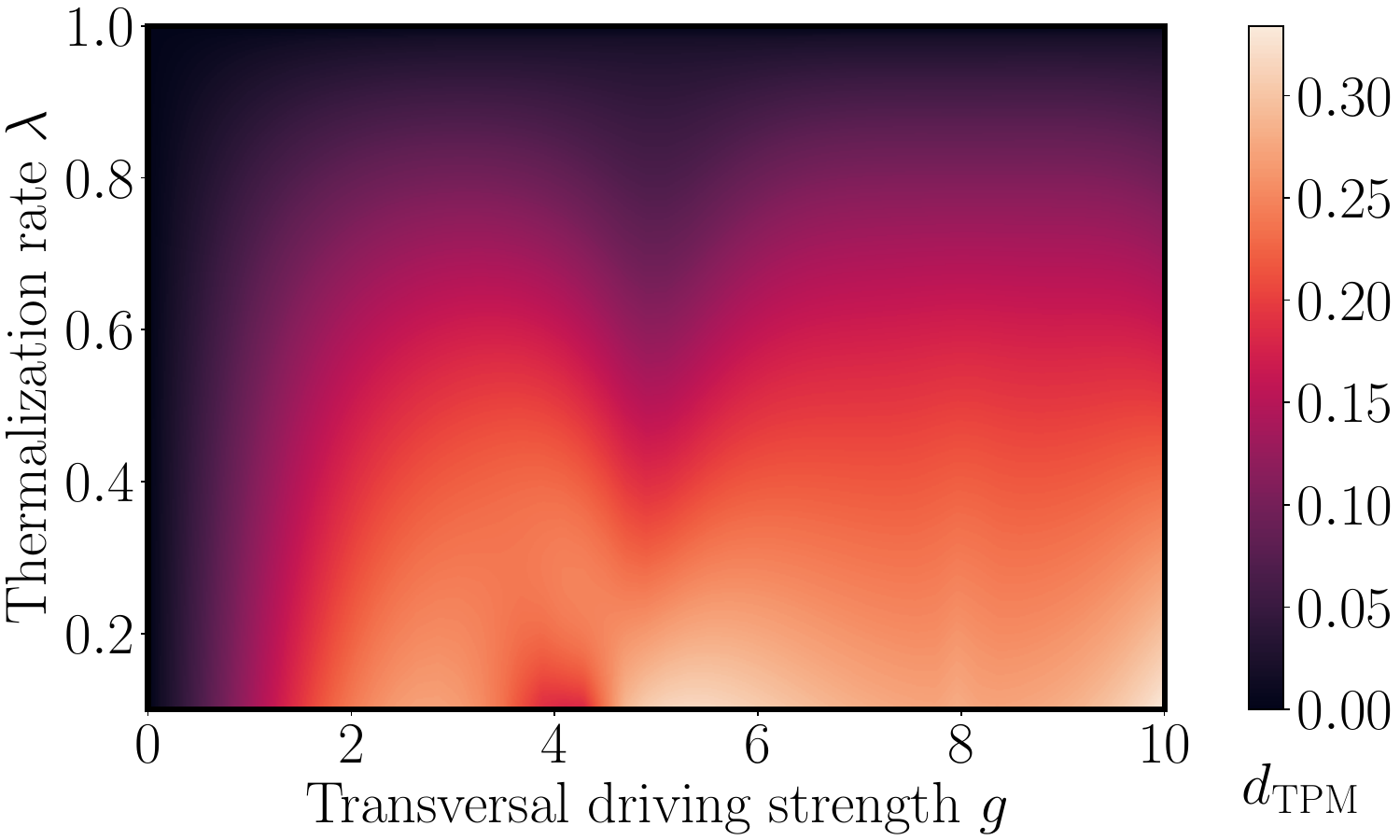}
    \caption{Trace distance $d(\langle \rho \rangle, \tilde \rho_1)$ between the average state $\langle \rho \rangle$ obtained after performing a round of successive TPM measurements starting and ending with the state after the cold isochore and the steady state $\tilde \rho_1$ of the unmeasured engine. Parameters  are $\omega_{\mathrm{h}} = 10$, $T_{\mathrm{h}} = 14$, $\omega_{\mathrm{c}} = 0.5$, $T_{\mathrm{c}} = 0.1$, $g = 9$, and $\lambda_{\mathrm{c}} = \lambda_{\mathrm{h}} = \lambda$.}
    \label{figsm:distance}
\end{figure}

In the main text, we further showed  that the DBN scheme is minimally invasive by evaluating the trace distance, $d(\langle \rho \rangle, \tilde \rho_1)$, between the averaged measured state $\langle \rho\rangle$ and the unmeasured steady state $\tilde \rho_1$. This trace distance vanishes for the DBN protocol but not for the TPM scheme, as seen in Fig.~2 as a function of the thermalization rate $\lambda$, but again for a fixed value of the transverse driving strength $g$. We here complement these findings with the contour plot of $d_\text{TPM}(\langle \rho \rangle, \tilde \rho_1)$ as $g$ is varied (Fig.~\ref{figsm:distance}).

\clearpage

\section{Violation of "universal" fluctuation bounds}
\begin{figure}[!htb]
    {%
        \includegraphics[width=0.48\linewidth]{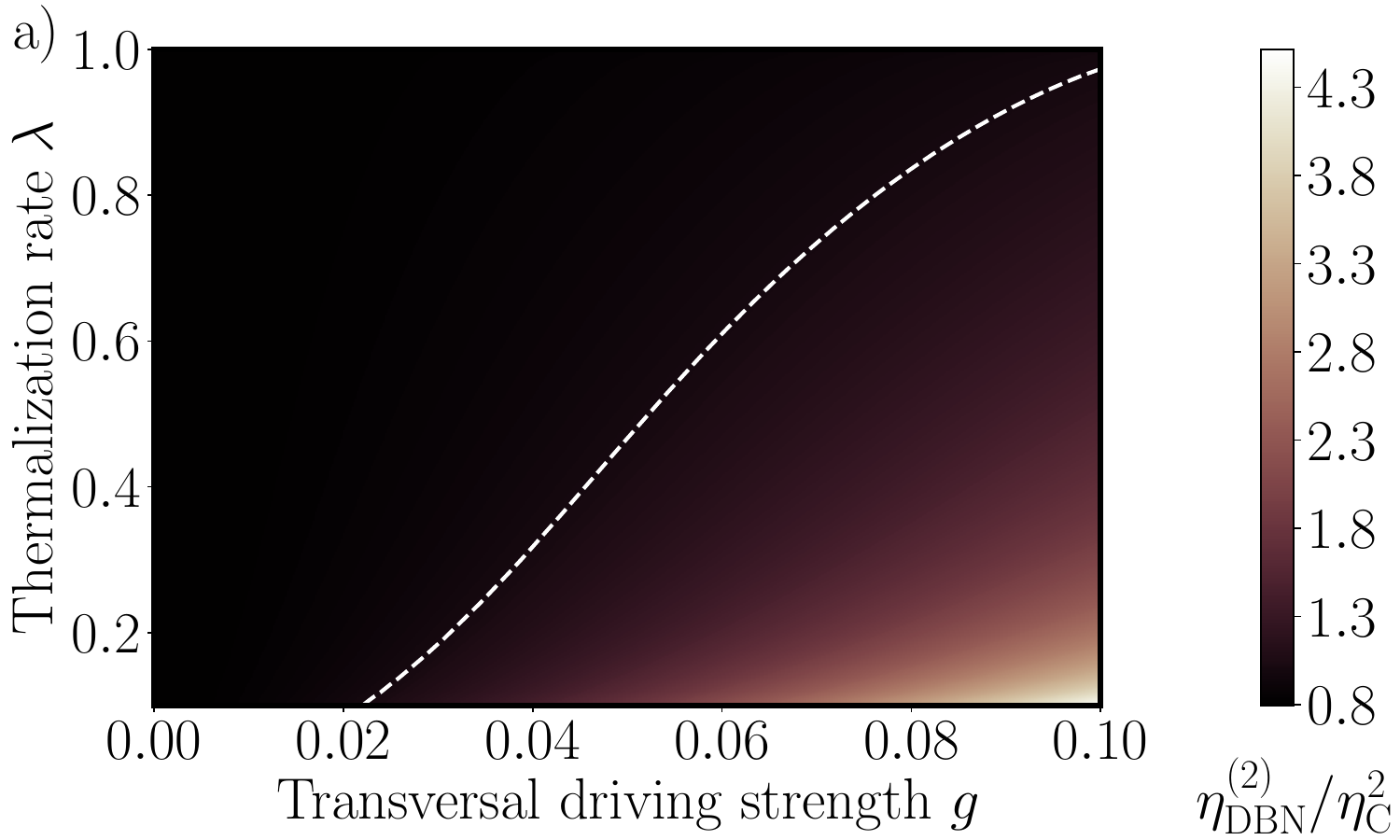}%
    } \hspace{0.02\linewidth}
     \label{figsm:boundcountourTPM}{%
        \includegraphics[width=0.48\linewidth]{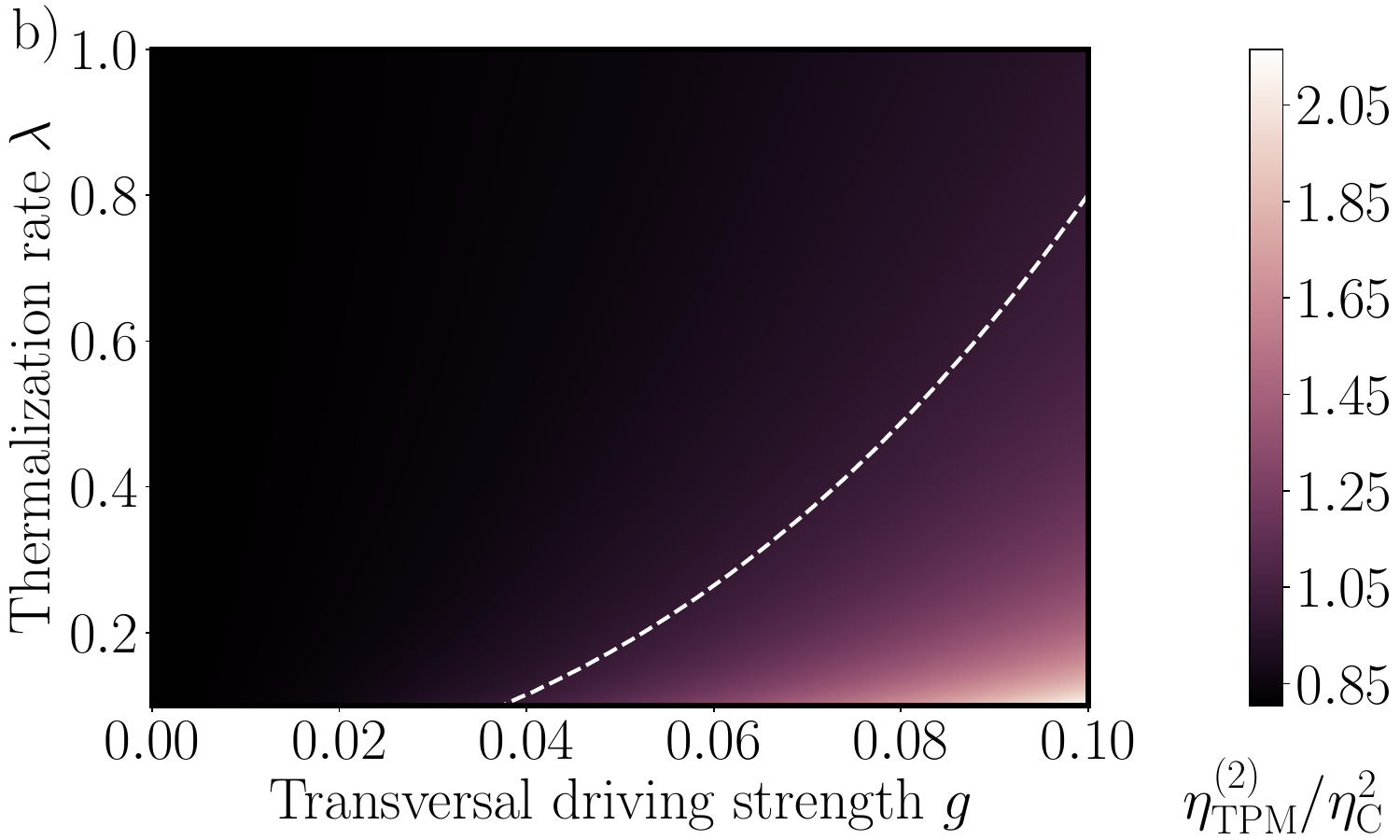}%
    }
    \caption{Quotient $\eta^{(2)} / \eta_{C}^{2}$ evaluated  with a) the DBN protocol and with b) the TPM scheme, in the quasistatic regime. The white dashed lines indicate the set of values such that $\eta^{(2)} / \eta_{C}^{2} = 1$. Every value above that line adheres to the bound, while everything below it violates it. Parameters used are $\omega_{\mathrm{h}} = 1$, $T_{\mathrm{h}} = 1.2$, $\omega_{\mathrm{c}} = 0.85$, $T_{\mathrm{c}} = 1$, and $\lambda_{\mathrm{c}} = \lambda_{\mathrm{h}} = \lambda$.}
    \label{figsm:boundcountour}
\end{figure}

\begin{figure}[!htb]
    \subfloat[Relative entropy of coherence $\mathcal{C}$ of the steady state before the isentropic compression step, $\tilde{\rho}_{1}$. \label{figsm:boundcoherence1}]{%
        \includegraphics[width=0.48\linewidth]{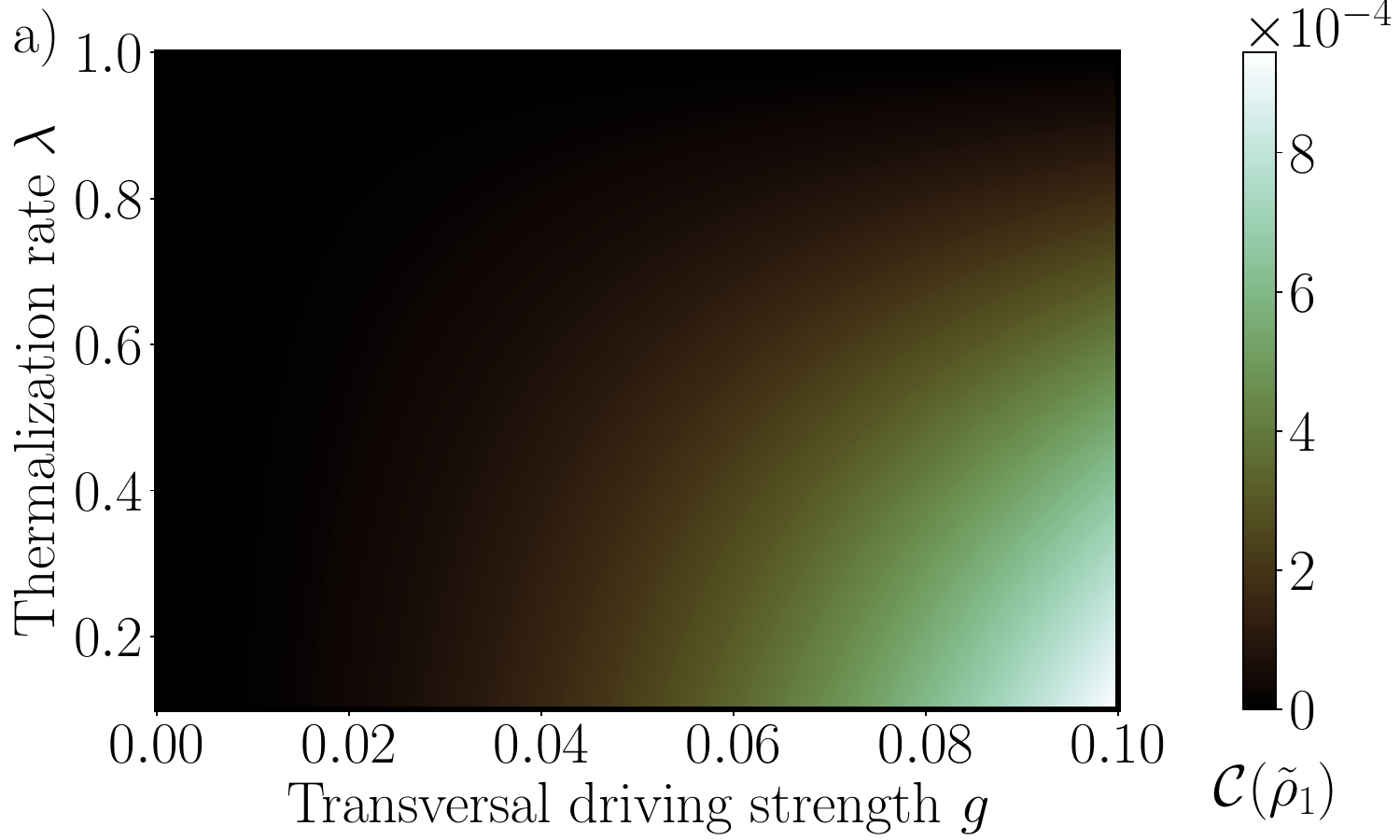}%
    } \hspace{0.02\linewidth}
    \subfloat[Relative entropy of coherence $\mathcal{C}$ of the steady state before the isentropic expansion step, $\tilde{\rho}_{3}$. \label{figsm:boundcoherence3}]{%
        \includegraphics[width=0.48\linewidth]{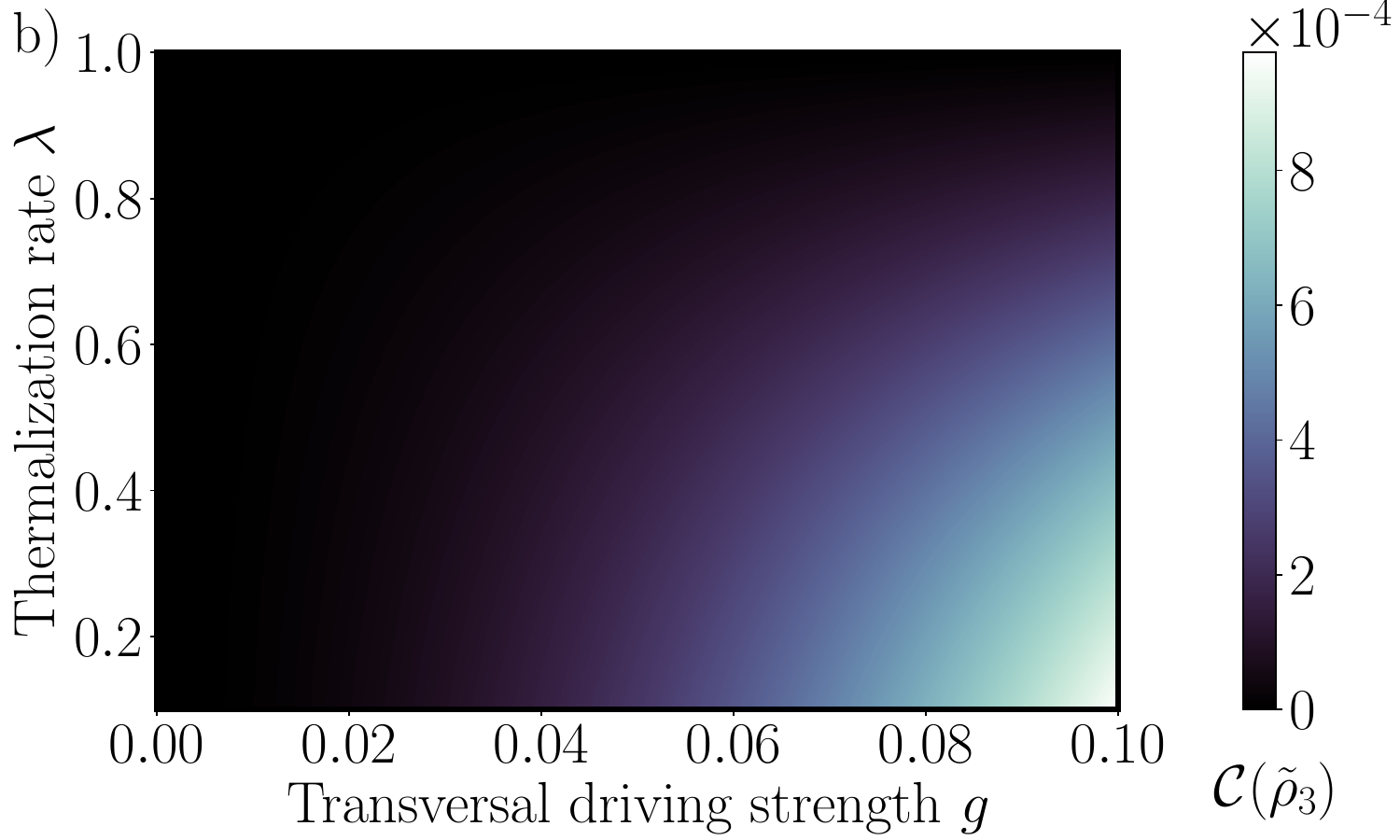}%
    }
    \caption{a) Relative entropy of coherence $\mathcal{C}$ of the steady state before the isentropic compression step, $\tilde{\rho}_{1}$, and b)  before the isentropic expansion step, $\tilde{\rho}_{3}$, in the quasistatic regime. Same parameters as in Fig.~\ref{figsm:boundcountour}.}
    \label{figsm:boundcoherence}
\end{figure}

Finally, we found that "universal" fluctuation bounds put forward for quasistatic engines \cite{agarwalla_2021_2s,agarwalla_2021s,das23s,moh23s,moh23as,watanabe_2025s} can actually be violated in coherent engines (Fig.~4 of the main text), for both measurement protocols. We complement this figure by additional contour plots for the ratio $\eta^{(2)} / \eta_{C}^{2}$, for varying transverse field strength $g$ (Fig.~\ref{figsm:boundcountour}). Figure \ref{figsm:boundcoherence} shows that these violations are related to the presence of coherence in the cycle (even though these are small in the quasistatic regime).

\begin{thebibliography}{99}
\bibitem{com} The fixed-point is unique since $\Lambda$ is a contractive CPTP map by construction (for $t_{\mathrm{h} , \mathrm{c}} > 0$) and the Hilbert space of $S$ is finite-dimensional \cite{rudin_1976, giovannetti_2007}.
\end{thebibliography}
\end{document}